\documentclass[12pt]{article}

\usepackage{amssymb,amsmath,amsfonts,eurosym,geometry,ulem,graphicx,caption,color,setspace,sectsty,comment,footmisc,natbib,pdflscape,subfigure,array,theorem, kpfonts, xcolor, tabularx, cellspace, setspace, booktabs, makecell, fancyhdr, pdflscape, adjustbox, multirow, rotating, placeins, tikz}

\usetikzlibrary{positioning,calc,arrows,arrows.meta}
\definecolor{fv}{HTML}{eff1f5}
\definecolor{mycyan}{HTML}{00bbdd}
\usetikzlibrary{decorations.pathreplacing,calligraphy}

\usepackage[toc,page]{appendix}
\usepackage{threeparttable}

\usepackage{floatrow}
\theorembodyfont{\rm}

\newcolumntype{L}[1]{>{\raggedright\let\newline\\arraybackslash\hspace{0pt}}m{#1}}
\newcolumntype{C}[1]{>{\centering\let\newline\\arraybackslash\hspace{0pt}}m{#1}}
\newcolumntype{R}[1]{>{\raggedleft\let\newline\\arraybackslash\hspace{0pt}}m{#1}}

\usepackage[utf8]{inputenc}
\usepackage [english]{babel}
\usepackage[babel=true]{csquotes}
\MakeOuterQuote{"}

\usepackage{natbib}
\usepackage{bbm}
\usepackage{comment}
\usepackage{enumerate}
\usepackage{nameref}

\newcommand{\A}{\textbf{A}}

\usepackage[hyphens]{url}
\usepackage[hidelinks]{hyperref}
\definecolor{softblue}{HTML}{1C46C0}
\definecolor{softred}{HTML}{B54747}
\definecolor{psered}{HTML}{DD0547}

\hypersetup{%
 colorlinks = true,
 linkcolor = psered,
 citecolor = psered,
 urlcolor = psered
}

\begin{document}

\tikzstyle{level 1}=[level distance=30mm, sibling distance=30mm]
\tikzstyle{level 2}=[level distance=30mm, sibling distance=15mm]
\tikzstyle{level 3}=[level distance=20mm]

\begin{titlepage}

\title{\fontsize{19pt}{19pt}\selectfont \sc{\bfseries A Two-Ball Ellsberg Paradox: An Experiment}}
\vspace{2cm}
\author{Brian Jabarian
\thanks{Paris School of Economics. Contact: \href{brian@jabarian.org}{brian@jabarian.org}.} 
\hspace{35pt} Simon Lazarus\thanks{Department of Economics, Princeton University.}}
\date{\today\footnote{We are indebted to Leeat Yariv and Jean-Marc Tallon for their continued guidance. We are grateful for the comments of Miguel Ballester, Roland Bénabou, Elias Bouacida, Marc Fleurbaey, Evan Friedman, Yoram Halevy, Nicolas Jacquemet, Yves Le Yaouanq, Pietro Ortoleva, Franz Ostrizek, Ryan Opra, P\"ellumb Reshidi, Evgenii Safonov, Elia Sartori, Denis Shishkin, and our fellow Ph.D. colleagues at the Princeton Department of Economics for their generous participation into our pilots. We are grateful for the research assistance provided by Christian Kontz and Andras Molnar. We thank seminar audiences at Princeton, MIT Sloan, and PSE. We thank the Princeton Experimental Laboratory for the Social Sciences for financial support. Princeton IRB approval \# 12380 was obtained on April 29, 2020.  This paper was partly written while Brian was visiting the Department of Economics at Princeton University and the Kahneman-Treisman Center for Behavioral Science and Public Policy in 2018-2020. He thanks their hospitality. He also acknowledges financial support from the Paris School of Economics, Sorbonne Economics Center, the Forethought Foundation, the Grant ANR-17-CE26-0003, and the Grant ANR-17-EURE-001 for his research visit to Princeton University.}}

\maketitle

\vspace{10pt}
\begin{center}
\textsc{\large \textcolor{black}{JOB MARKET PAPER\footnote{This is Brian Jabarian's job market paper.}}\\[10pt]
\href{https://arxiv.org/pdf/2206.04605.pdf}{link to the most recent version}
}
\end{center}
\vspace{10pt}

\begin{abstract}
\onehalfspacing
\fontsize{10pt}{12pt}\selectfont
\noindent 

We conduct an incentivized experiment on a nationally representative US sample \\ (N=708) to test whether people prefer to avoid ambiguity even when it means choosing dominated options.  In contrast to the literature, we find that 55\% of subjects prefer a risky act to an ambiguous act that always provides a larger probability of winning.  Our experimental design shows that such a preference is not mainly due to a lack of understanding.  We conclude that subjects avoid ambiguity \textit{per se} rather than avoiding ambiguity because it may yield a worse outcome. Such behavior cannot be reconciled with existing models of ambiguity aversion in a straightforward manner. 

\vspace{0in}

\bigskip
\end{abstract}

\setcounter{page}{0}
\thispagestyle{empty}
\end{titlepage}
\pagebreak \newpage

\onehalfspacing

\section{Introduction}\label{sec:Introduction}

We present in this paper a novel and simple experiment that documents a new channel through which decision-makers might shy away from ambiguous choice situations and prefer risky situations that are ``objectively" worse for them. \textit{Risk} refers to situations where various outcomes can occur with precise probabilities. \textit{Ambiguity} (or uncertainty) refers to situations where agents cannot attach precise probabilities to the various possible outcomes. \textit{Ambiguity aversion} refers to the preference for risky situations over similar ambiguous situations.

In 1961, Daniel Ellsberg (\cite{Ellsberg}) showed in his famous ‘Ellsberg Paradox’ experiment that people generally exhibit ambiguity aversion, falsifying the popular Subjective Expected Utility theory of decision-making under uncertainty (\cite{savage1972foundations}). Since then, theorists have devised many other models of rational decision-making that account for the Ellsberg Paradox. Amid the early axiomatic work, \cite{Schmeidler} proposed Choquet expected utility (CEU), \cite{GilboaSchmeidler} designed Maximin expected utility (MEU), \cite{GMM} suggested the alpha-maximin expected utility ($\alpha$-MEU). The second generation of models includes \cite{Klibanoff}, \cite{MMR}, and \cite{Strzalecki}. All these models can accommodate typical Ellsberg choices, but they cannot explain the results of our experiment.

Ambiguity and ambiguity aversion are relevant to economists and policymakers since, in most real-world situations, agents cannot attach precise probabilities to the possible outcomes. Relying on the standard models cited above, researchers have explored the important implications of ambiguity in diverse economic fields.  To name a few: in environmental economics, \cite{millner2013scientific} demonstrate the effects of ambiguity on the social cost of carbon by integrating ambiguity within Nordhaus' famous integrated assessment model of climate economy.  \cite{lange2008uncertainty} investigate the learning effects of climate policy under ambiguity. In health economics, \cite{treich2010value} shows under which conditions ambiguity aversion increases the value of a statistical life. In macro-finance,  \cite{ju2012ambiguity} show how ambiguity aversion can account for the equity premium puzzle. 

Our paper focuses on the following thought experiment proposed by \cite{Jabarian1}. There are two urns, \textbf{R} and \textbf{A}, and two types of balls: red and blue. In urn \textbf{R}, there are 100 balls: 50 red and 50 blue balls. In urn \textbf{A}, there are 100 balls in unknown proportion. Subjects draw two balls successively (with replacement) from one of the two urns and win a prize if the balls match in color. We focus on the following gambles. $RR$: drawing twice from urn \textbf{R}, versus $AA$: drawing twice from urn \textbf{A}. Which gamble would subjects prefer: $RR$ or $AA$? 

Although urn \textbf{R} may seem more attractive than urn \textbf{A}, since its content is merely \textit{risky}, gamble $AA$ has an unambiguously \textit{larger} win probability than gamble $RR$. Indeed, the more \emph{unevenly} distributed urn \textbf{A}'s contents, the higher the win probability of gamble $AA$. Urn \textbf{A} having 50\% red balls and 50\% blue balls is the \textit{worst case}: it yields a probability of winning equal to $\frac{1}{2}$. Any other distribution gives a probability of winning that is higher. If the distribution were, say, $(\frac{1}{3},\frac{2}{3})$, the probability of winning would be $\frac{5}{9}>\frac{1}{2}$. We call the preference for gamble $RR$ over $AA$ the \textit{Two-Ball Ellsberg Paradox}, and among a nationally representative sample in an incentivized online experiment, we find that it is widespread.

Unless subjects have beliefs over the two draws that are not consistent with the information given to them (say, subjects somehow believe the draws are not independent), the choice of \textit{RR} over \textit{AA} cannot be reconciled with existing models of ambiguity aversion in a straightforward manner. For instance, in the Maxmin Expected utility model of \cite{GilboaSchmeidler}, if subjects entertained the entire simplex over $\{R,B\}$ for the composition of an urn \textbf{U} and form beliefs over the two draws by composing each prior with itself, this would lead to indifference between gambles $RR$ and $AA$.\footnote{Dominance, as in \cite{gajdos2008attitude}, implies that $RR$ cannot be strictly preferred to $AA$.} 
Actually, if one assumes that the set of priors is a subset of $\{q \in \Delta(\{R,B\}^2) | q=p\times p; p\in\Delta(\{R,B\})\}$, then the choice \textit{RR} over \textit{AA} is incompatible with $\alpha$-maxmin expected utility for any $\alpha$. This choice is also incompatible with \cite{savage1972foundations}'s Subjective Expected Utility model if beliefs are a product measure of the type $p\times p$.

\paragraph{Aims of the paper.} We report the results of an incentivized experiment on a nationally representative US sample to test whether people show Two-Ball Ellsberg Paradox preferences and if so, to understand why this may be the case.  Our experiment aims to answer two main questions. First, do subjects occasionally avoid an ambiguous act by choosing an unambiguous lottery worse than any lottery the ambiguous act could produce? Second, is this preference entirely due to failing to understand Two-Ball gambles?  We also investigate how subjects' Two-Ball Ellsberg Paradox preferences relate to other ``paradoxical'' behaviors, such as ambiguity aversion (as in the Classical Ellsberg Paradox) and complexity aversion (as measured by two-ball versus one-ball gambles, by compound versus simple gambles by the combination of the two).

\paragraph{Experimental design.}
 We used a multiple price list (MPL) to elicit subjects' Certainty Equivalents (CEs), allowing us to determine the strength of their preferences for various gambles. Laboratory, and even more online, experiments eliciting subjects' CEs for gambles are often subject to significant measurement error, i.e., biases in estimating coefficients and correlations. We rely on the \textit{Obviously Related Instrumental Variables} (ORIV) approach of \cite{GSY} to correct these errors.

Due to time constraints, our experiment is divided into four treatments and subjects are randomly assigned to one of them. Treatment \textsc{\bfseries nudging} tests whether subjects' preference for urn \textbf{R} over urn \textbf{A} is due to their failing to understand the gambles.  Treatment \textsc{\bfseries complexity} tests whether subjects respond in the same way to \textit{complexity}, in the form of compound lotteries, as they do to ambiguity.  Treatment \textsc{\bfseries robustness} tests whether subjects' preference for urn \textbf{R} over urn \textbf{A} is due to a false belief that the contents of urn \textbf{U} are changed between draws from that urn, and it also explores how the \textit{number or proportion of draws that are from urn} \textbf{U} affects subjects' preferences.  Lastly, treatment \textsc{\bfseries order} tests whether the order of draws from various urns matters and explores whether the \textit{mere presence of ambiguity} affects subjects' decisions.

\paragraph{Main results.}
Overall, we determined that 55\% of subjects exhibit a distaste for ambiguity that cannot be explained by most existing models or by the fact that subjects failed to understand the situations presented to them.  Our results suggest that subjects avoid ambiguity \textit{per se} as opposed to avoiding it simply because it may yield a worse outcome. 
The two most important findings supporting this idea of ambiguity aversion \textit{per se} originate from our core treatment, \textsc{\bfseries nudging}. 

This treatment revealed that a lack of understanding does not entirely explain the preference for avoiding ambiguity. Subjects ``correctly'' select the act with a larger win probability when comparing two similar two-ball acts that are \textit{both} ambiguous; that is, they correctly identify that more unevenly distributed urns are preferable in two-ball gambles.  Furthermore, those same subjects went on to exhibit as strong a preference for $RR$ over $AA$ as subjects who were not exposed to such an opportunity to be ``nudged'' toward identifying more uneven urns as better.  This suggests that subjects' preference for avoiding ambiguity -- even when it can only improve their odds of receiving a given prize -- cannot be due to failing to understand the gambles.

\paragraph{Contribution to the literature.}

The recent literature has produced several experimental tests and thought experiments that challenge existing decision-making models under risk and ambiguity. The following are the most resembling the one in this paper, either because they present some scenarios to decision-makers that resemble our two-ball gamble or because they reach similar conclusions to ours. 

\cite{epstein2019ambiguous} employ a similar two-ball gamble in one of their supplemental treatments from a 2014 experiment: subjects win the gamble if and only if two balls drawn with replacement from a single ambiguous urn (containing at most two different colors of balls) have the same color.  However, they do not elicit subjects' Certainty Equivalents for this gamble and, crucially, do not observe the choice over a risky bet; they merely ask whether subjects prefer this gamble or a gamble that wins if and only if the first of the balls drawn is red (essentially, a 1-ball ambiguous gamble).  Hence, from their results, one cannot directly estimate the proportion of people who prefer a 50-50 risky gamble to a two-ball ambiguous gamble.  However, among subjects in their experiment who exhibited monotone and transitive choices, 21.6\% exhibited a strict preference for the 1-ball ambiguous gamble over the 2-ball ambiguous gamble - a proportion that is consistent with our results when one takes into account the possible preference for a 50-50 risky gamble over a 1-ball ambiguous gamble.

\cite{fleurbaeyambiguity2019} constructs a thought experiment in which there is a risky urn $R$ containing 50 red and 50 blue balls and an ambiguous urn $A$ containing only red and blue balls; the decision maker draws two balls sequentially from some combination of these urns and wins if the two balls have the same color.\footnote{We thank Marc Fleurbaey for having provided access to his unpublished manuscript.}  The decision maker must choose to either (1) draw a ball from urn $R$, observe its color, then choose to draw the second ball from either (a) urn $R$ (with replacement) or (b) urn $A$; or (2) draw a ball from urn $A$ and then a ball from urn $R$.  The payoffs from winning vary slightly between options (1a), (1b), and (2) in such a way as to generate time-inconsistent or otherwise ``unreasonable" behavior from certain theoretical types of decision-makers.  In contrast to this thought experiment, our paper's central two-ball gamble compares two draws from urn $A$ to two draws from urn $R$, and the decision maker does not have any choices between the drawing of the first and second balls.  Although both papers consider situations in which individuals may pay to avoid the mere presence of ambiguity, only our Two-Ball Ellsberg Paradox is an example in which individuals choose a dominated gamble to avoid such ambiguity.

\cite{yang2017testing} designed an experiment in which two balls are drawn with replacement from a single urn containing only red and white balls; the subject wins 50 yuan if the first ball is red and independently wins an additional 50 yuan if the second ball is white.  This gamble has the properties that (i) its expected payout is 50 yuan regardless of the composition of the urn and (ii) the variance of this payout \textit{decreases} as the urn's contents become more dispersed (e.g., it has zero variance when the urn is all red balls or all white balls).  Thus, when choosing whether to play this gamble drawing with replacement from a risky urn $R$ whose contents are known to be 50\% each color versus an ambiguous urn $A$ whose distribution between red/white is unknown, a decision maker with risk-averse decision maker should think urn $A$ is better than urn $R$, assuming her preferences satisfy the monotonicity axiom.  After eliciting subjects' risk attitudes, the authors estimate that as many as 45\% of risk-averse subjects choose urn $A$ over urn $R$, violating all theories that include a monotonicity axiom.  These results are qualitatively similar to the ones we find in our present paper, except that our central two-ball gamble's payoff has not only \textit{variance} that decreases with the dispersion of the urn's contents but also has \textit{mean} that increases with this dispersion.  Hence, urn $A$ may be attractive to those who are risk-averse and those who are risk-seeking.

More broadly, our paper belongs to the literature on experimental tests and thought experiments questioning models of decision-making under risk and ambiguity. Among the most famous and relevant experiments, we can cite the following.
\cite{Machina2009} presents thought experiments demonstrating plausible violations of the CEU model.  \cite{MachinaParadoxes} employ variations on \cite{Machina2009} to propose plausible violations of several of the decision-making models mentioned above. \cite{ReflectionExperiment} test one of these examples, the so-called "reflection example" of \cite{Machina2009}, in an experimental setting and reject the MEU and variational preferences models and the smooth ambiguity model of \cite{Klibanoff}. \cite{Blavatskyy} proposes a variant of Machina's reflection example, casting doubt on more recent decision-making models under ambiguity, such as \cite{VectorEU}'s \textit{Vector Expected Utility}. 

\cite{Halevy} demonstrates that even in relatively simple settings, individuals' choices may be inconsistent with models that assume that individuals correctly reduce compound lotteries - an assumption that is implicitly part of models built on the framework of \cite{AnscombeAumann}. He further shows that ambiguity aversion in the classic experiment of \cite{Ellsberg} is closely correlated with an adverse reaction to compound lotteries. \cite{GSY} replicate \cite{Halevy}'s experiment with a correction for measurement error, revealing that this correlation may be close to 1. \cite{Schneider} examine whether subjects' preferences satisfy \textit{weak separability}, a weak form of the monotonicity axiom of \cite{AnscombeAumann}, and find that nearly half of the subjects violate it. Furthermore, subjects violating the monotonicity axiom generally make choices consistent with first-order stochastic dominance when choosing under risk, demonstrating that these violations are likely unrelated to a lack of understanding of their choices. 

Our experiment distinguishes itself from these by proposing a new class of decision problems: Two-Ball Ellsberg drawings versus non-ambiguous drawings. As the Two-Ball Ellsberg drawings feature ambiguity but guarantee a minimum win probability at least as large as a related non-ambiguous gamble, they enable us to test whether a subject avoids ambiguity \textit{per se} instead of avoiding ambiguity because it may yield a worse outcome.

\paragraph{Structure of the paper.} Section \ref{sec:experimentaldesign} provides an overview of our experimental design and methodology.  Section \ref{sec:data} presents the results of our core gambles: the Two-Ball Ellsberg Paradox.  Sections \ref{sec:nudging}, \ref{sec:complexityaversion}, \ref{sec:robustness} and \ref{sec:distaste} describe and summarize the results of our experimental treatments meant to test various hypotheses about what may be generating the Two-Ball Ellsberg Paradox.  Section \ref{sec:discussion} discusses some remaining plausible explanations of subjects' behavior, and section \ref{sec:conclusion} concludes.

\section{Experimental Design}\label{sec:experimentaldesign}

Our experiment was designed to answer two primary questions. First, to what extent do subjects prefer urn \textbf{R} over urn \textbf{A} in our two-ball gamble? Second, what possible explanations of this ``paradoxical'' preference can be falsified? Answering the first question only requires asking subjects about a few different gambles.  However, since many possible explanations exist for a preference for urn \textbf{R} over urn \textbf{A}, our experiment includes many gambles designed to address the second question.

We used Prolific to run our experiment and collect our data. Prolific is an online survey platform that, due to its participant pool's quality, is increasingly used in economics to conduct surveys and incentivized experiments.  Our sample comprised 880 participants, selected to be nationally representative in age and gender. Of these initial 880 participants, 708 passed the basic attention-screening questions and criteria described at the end of this section.

Due to the constraints on subjects' time and attention inherent in an online experiment, our various gambles were divided across four treatments, with each subject completing exactly one treatment.  All treatments ask subjects about our central two-ball gamble (playing with urn \textbf{A} versus urn \textbf{R}), and all treatments elicit subjects' ambiguity attitudes via the classic two-urns Ellsberg paradox.  Beyond this, each treatment contains some gambles specific to that treatment.  Similar gambles were grouped into \textit{blocks}, and gambles within a block were presented in random order.\footnote{The order of the blocks was also randomized; we detail the particular randomization for each treatment as elaborated in the sections \ref{sec:nudging}, \ref{sec:distaste}, \ref{sec:complexityaversion} and \ref{sec:robustness}.}

In each gamble, the subject can either "win" (gain \$3) or "lose" (gain nothing). After viewing instructions explaining the conditions under which the current gamble will win or lose, the subject must report her certainty equivalent (CE) for that gamble from a multiple price list (MPL) containing dollar amounts between \$0 and \$3 in increments of 10 cents. Compared to eliciting choices, the MPL allows us to measure the intensity of subjects' preferences.

Laboratory and online experiments eliciting subjects' CEs for gambles are often prone to significant measurement error.  To correct this, we rely on the \textit{Obviously Related Instrumental Variables} (ORIV) method of \cite{GSY}. Compared to other methods to correct measurement errors, such as using the first elicitation as an instrument for the second, the ORIV approach generally results in lower standard errors.  We, therefore, elicit subjects' CEs \textit{twice} for most of our gambles.

Including all duplicate questions, each treatment contains 11 or 12 gambles in total.  In each treatment, three\footnote{Although incentivizing only \textit{one} gamble would allow us to raise the monetary stakes of each question, doing so would create too large a variance in different subjects' payoffs, which was undesirable for this online experiment.} of these gambles were selected at random for incentivization: if a gamble was selected, then a random row of the MPL for that gamble was chosen, and subjects were given what they reported they preferred from that row.\footnote{For example, if Gamble X was selected for incentivization, and then the row ``\$1.20'' was selected at random for this gamble, the following happens.  (A) If the subject reported she preferred a fixed \$1.20 payment to play Gamble X, then she received \$1.20.  (B) If the subject reported she preferred playing Gamble X to receiving \$1.20, then we simulated Gamble X and gave her \$3 if it won and \$0 if it lost.}  Subjects received an average payment of \$3.50 from the incentivized questions, plus a fixed \$2 payment for completing the experiment.

Since the monetary stakes of the experiment were not very high, there is a reason for concern that subjects may answer at random to quickly finish the experiment.  We employed three screening criteria to address this concern: (1) After the experiment instructions, but before the gambles, subjects were given a 3-question basic comprehension quiz about the instructions.  Any subject who failed at least one of these questions was given a small payment and forced to leave the experiment.  (2) Subjects were given a standard attention-screening question between each of the experiment's major sections.  Subjects failing at least one such question were removed from our analysis.  (3) If, across our two elicitations of a subject's CE for the same gamble, the subject reported two CEs that differed by more than \$1 (that is, one-third the size of the \$3 MPL table), that subject was removed from our analysis.\footnote{Other reasonable thresholds for exclusion, such as ``differed by more than \$1.50,'' yield qualitatively similar results in our analysis as detailed in the Appendix.}  Out of an initial pool of 880 subjects, 172 were removed due to violating at least one of the criteria (1)-(3).

\section{Main Result: Participants Fall for the Two-Ball Ellsberg Paradox}\label{sec:data}

The block \textbf{2Ball} contains this experiment's central gambles and is present in all four of our treatments.  It contains two gambles, named $RR$ and $AA$. The block \textbf{Ellsberg} replicates the classic Ellsberg paradox to elicit subjects' attitudes towards risk and ambiguity and is also present in all four of our treatments; it contains two gambles named $R$ and $A$.  Table \ref{table:2BallEllsberg} describes these gambles.

\bigskip
\begin{table}[H]
\begin{center}
    \begin{tabular}{cl}
    \toprule \toprule
        \textsc{gamble name} & \multicolumn{1}{c}{\textsc{gamble description}} \\
        \toprule
        \multicolumn{2}{c}{Block \textbf{2Ball}} \\
         $RR$ &  Draw 2 balls with replacement from urn \textbf{R} = [50 red, 50 blue]; \\
         & win if the two balls have the same color. \\
       \midrule
         $AA$ & Draw 2 balls with replacement from urn \textbf{A} = [Unknown red, Unknown blue]; \\
         & win if the two balls have the same color. \\
         \hline
         & \\
        \toprule
        \multicolumn{2}{c}{Block \textbf{Ellsberg}} \\
         $R$ &  Choose a color.  Draw a ball from urn \textbf{R} = [50 red, 50 blue]; \\
         & win if the drawn ball has the color you chose. \\
         \midrule
         $A$ &  Choose a color.  Draw a ball from urn \textbf{A} = [Unknown red, Unknown blue]; \\
         & win if the drawn ball has the color you chose. \\
         \bottomrule \bottomrule
 \end{tabular}
\caption{\sc {\bfseries description of gambles present in all treatments}}
\label{table:2BallEllsberg}
\end{center}
\end{table}
\bigskip

The blocks \textbf{2BallD} and \textbf{EllsbergD} contain duplicate gambles of those in blocks \textbf{2Ball} and \textbf{Ellsberg}. The standard practice when double-eliciting CEs requires the two ``duplicate'' gambles measuring the same CE to have slightly different wordings so as two constitute two \textit{independent} measurements of that CE.  To accomplish this, whenever we duplicate a block of gambles, we slightly change the specified \textit{total} number of balls in a given urn without changing the \textit{proportion} of balls of each color.  For example, in block \textbf{2BallD}, urn \textbf{R} contains 40 red and 40 blue balls rather than 50 red and 50 blue.

\begin{figure}[H]
\begin{center}
\includegraphics[scale = 0.4]{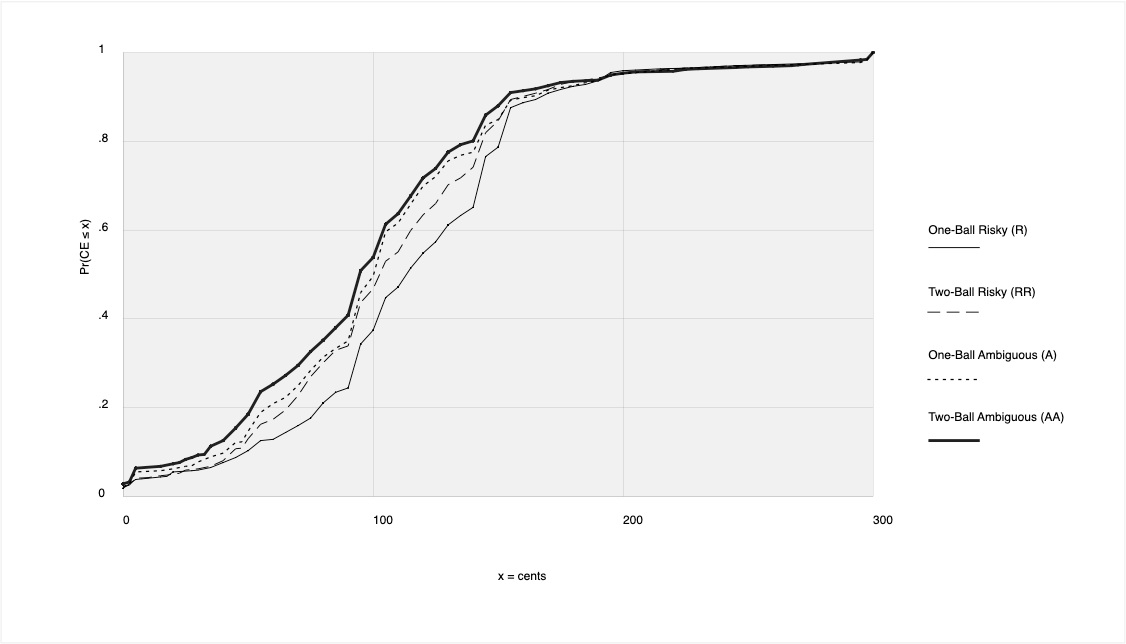}
\end{center}
\caption{\sc {\bfseries Cumulative Distribution of CEs for R, A, RR, AA}}
\label{2EllsbergCDFs}
\end{figure}

For each gamble $X$ that is double-elicited, we use the notation $X_i^j$ to represent the $j$-th elicitation of subject $i$'s CE for gamble $X$, and we use the notation
\[ X_i = \frac{X_i^1 + X_i^2}{2} \]
to denote the \textit{average} CE of subject $i$ for gamble $X$.  So, for example, $RR_{36}^2$ represents the 2nd elicitation of subject 36's CE for gamble $RR$, and $A_{15}$ denotes subject 15's \textit{average} CE for gamble $A$. Figure \ref{2EllsbergCDFs} shows the CDFs of the empirical distributions of the CEs for $RR$, $AA$, $R$, and $A$; Table \ref{tab:rawvariables} in Appendix \ref{sec:appendixraw} contains summary statistics for each elicitation of these CEs.

Other than at a few extreme CE values that were reported by a total of less than 10\% of subjects, these empirical CDFs lie in the same vertical order everywhere.  This suggests that on average, subjects prefer the gambles in the order
\[ R \succ RR \succ A \succ AA. \]

\begin{figure}[H]
\begin{center}
\includegraphics[scale = 0.72]{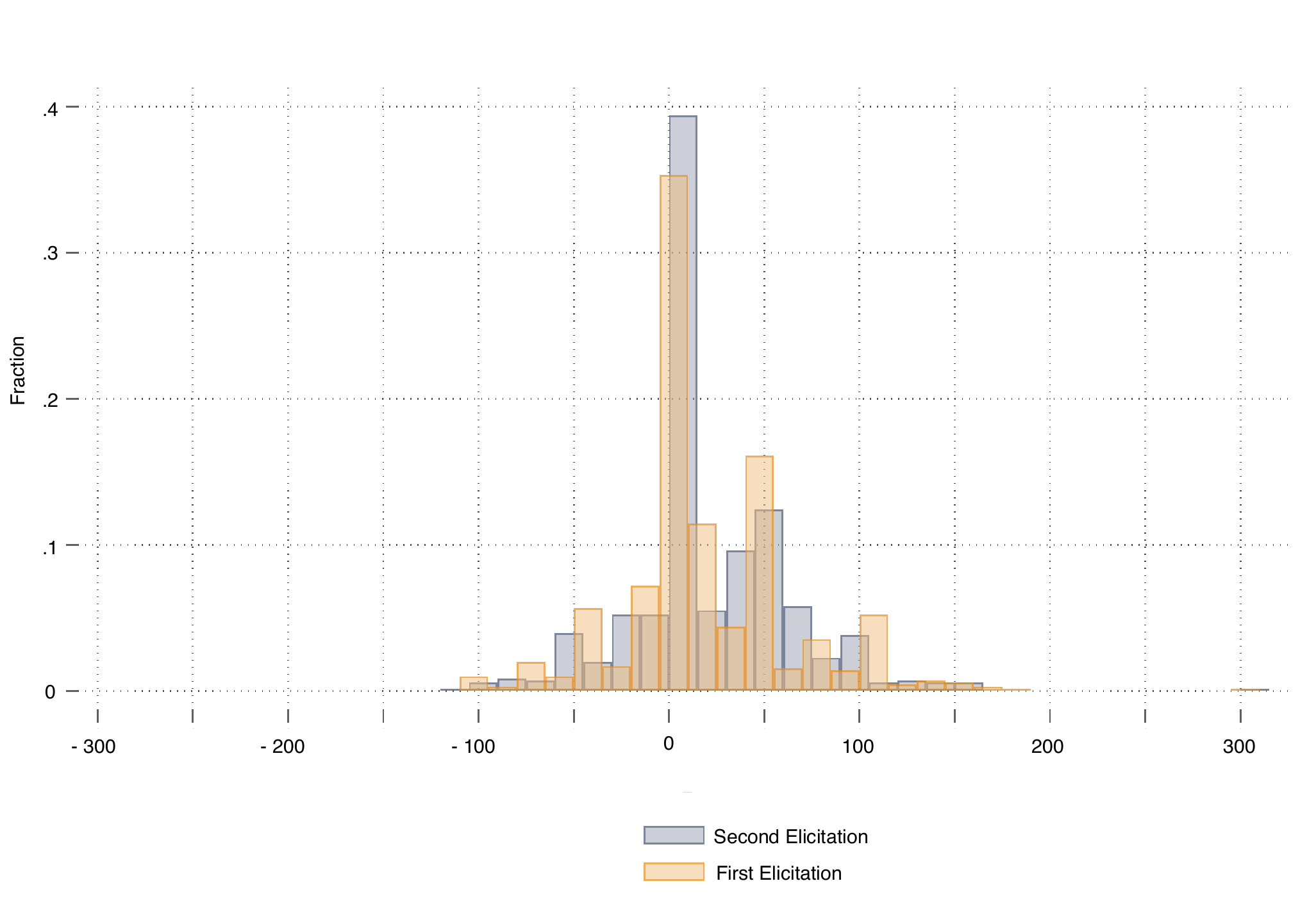}
\end{center}
\caption{\sc {\bfseries Histogram of $R - AA$, by elicitation}}
\label{ZHistogram}
\end{figure}

Nearly all widely-used models of decision making under risk and ambiguity cannot explain a preference for $R$ over $AA$ \textit{or} a preference for $RR$ over $AA$, since gamble $AA$ has a win probability of \textit{at least} 50\% while gambles $R$ and $RR$ have a win probability of \textit{exactly} 50\%.  Throughout this paper, we use the variable $R-AA$ to measure the extent to which individuals exhibit the ``2-Ball Ellsberg Paradox.''  Although it may be slightly more natural to compare gamble $AA$ to gamble $RR$, we choose to compare it to gamble $R$ since gamble $R$ provides a more standard basis of comparison for the other gambles in our experiment, such as the compound lottery $C$ in treatment \textsc{\bfseries complexity}.\footnote{See Section \ref{sec:complexityaversion} below.}  In any case, among the standard models of decision making, $R-AA$ taking a statistically significant positive value is sufficient to falsify all the same models would be falsified by a statistically significant positive value of $RR-AA$.  Figure \ref{ZHistogram} shows the distribution of individuals' reported CE differences $R^j-AA^j$ in each of the two elicitations $j$.


Averaging across both elicitations, a majority (54.9\%) of subjects exhibited 2-Ball Ellsberg Paradox preferences by reporting a value $R-AA$ that was greater than zero.  The average CE for gamble $R$ is 118.13 cents, while the average CE for gamble $AA$ is only 101.03 cents.  The 17.1 cent difference between these averages is statistically significant $(t = 11.7)$; individuals are willing to pay about 17\% more for gamble $R$ than they are for the higher-win-probability gamble $AA$.


\section{Is The Two-Ball Ellsberg Paradox Due to a Lack of Understanding?}\label{sec:nudging}

One natural explanation for subjects' preference for gamble $R$ over gamble $AA$ is that subjects \textit{fail to understand} the fact that gamble $AA$ must have \textit{at least as large} of a win probability than gamble $R$.  Our treatment \textsc{\bfseries nudging} was designed to test this hypothesis in two ways.  First, we present subjects with a variation of the 2-ball ambiguous gamble wherein we place \textit{bounds} on the contents of urn \textbf{A} to test if subjects can successfully identify that more \textit{unevenly} distributed urns yield higher win probabilities in 2-ball gambles.  We find that subjects \textit{do} correctly identify this -- or at least, they make choices consistent with such an understanding.  Second, we check if those subjects who were exposed to Treatment \textsc{\bfseries nudging} -- who "correctly" identified more unevenly distributed urns as more preferable -- tended to exhibit less of a preference for gamble $R$ over gamble $AA$ than those who were not exposed to this treatment.  We find that subjects in Treatment \textsc{\bfseries nudging} do not show a statistically significant difference in their values $R-AA$ from subjects not exposed to this treatment ($p=.63$).

\subsection{Description of Treatment \textsc{\bfseries nudging}}
The gambles unique to treatment \textsc{\bfseries nudging} are those in block \textbf{BoundedA}.  In this block, subjects play a 2-ball gamble: two balls are drawn from an urn \textbf{A} containing 100 balls, all red or blue, but whose exact contents are unknown. The subject wins \$3 if the two balls have the same color.  In each gamble in block \textbf{BoundedA}, some further information is given about the contents of urn \textbf{A}, as described in Table \ref{table:gambles_nudging}.
\bigskip
\begin{table}[H]
\begin{center}
    \begin{tabular}{|c|l|}
    \hline 
        \textsc{gamble name} & \multicolumn{1}{c|}{\textsc{gamble description}} \\
        \cmidrule{1-2}
        \multicolumn{2}{|c|}{Block \textbf{BoundedA}} \\
        \hline
        $BB^{40-60}$ & Urn \A{} is known to contain between 40 and 60 red balls. \\
        \hline
         $BB^{60-100}$ & Urn \A{} is known to contain between 60 and 100 red balls. \\
        \hline
         $BB^{95-100}$ & Urn \A{} is known to contain between 95 and 100 red balls. \\
         \hline
 \end{tabular}
\caption{\sc {\bfseries description of gambles present only in treatment nudging}}
\label{table:gambles_nudging}
\end{center}
\end{table}
\bigskip

\begin{figure}[H]
\resizebox{15cm}{!}{
\begin{tikzpicture}
\tikzstyle{block}=[draw, outer sep=0pt, inner sep=0pt, minimum width=4cm, minimum height=1cm, rounded corners=.5mm]
\node[block,minimum width=2.5cm, fill=gray!20] (a) {BoundedA};
\node[block, below=5mm, anchor=north west] (b) at (a.south west) {Ellsberg, 2Ball};
\node[block, right=8mm of a] (c) {Ellsberg, 2Ball};
\node[block, right=8mm of c] (d) {EllsbergD, 2BallD};
\node[block,minimum width=2.5cm, below=5mm of c, anchor=north east,fill=gray!20] (f) at (d.south east) {BoundedA};
\node[block] (e) at ($(b.east)!.5!(f.west)$) {EllsbergD, 2BallD};
\coordinate (n) at ([xshift=-3cm]$(a)!.5!(b)$);
\draw[rounded corners=2mm] (n)--node[left, pos=0]{\textsc{\bfseries nudging}}++(.7,0)|-(a.west);
\draw[rounded corners=2mm] (n)--++(.7,0)|-(b.west);
\draw (a)--(c) (e)--(f);
\draw[dotted, thick]  (c)--(d) (b)--(e);
\end{tikzpicture}}
\caption{\sc {\bfseries structure of treatment nudging}}
\label{fig:structure_nudging}
\end{figure}
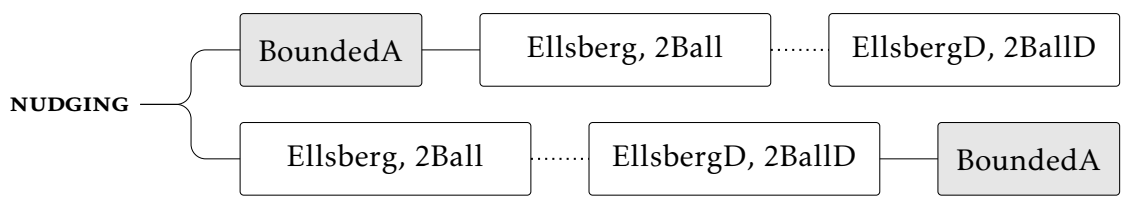

In treatment \textsc{\bfseries nudging}, subjects complete the blocks \textbf{BoundedA}, \textbf{Ellsberg} and \textbf{2Ball} as well as the duplicate blocks \textbf{EllsbergD} and \textbf{2BallD}.  The order in which these blocks were presented was determined at random, independently for each subject assigned to this treatment, according to Figure \ref{fig:structure_nudging}.  

\bigskip
\bigskip

In this figure, the initial split between line 1 (with BoundedA at the beginning) and line 2 (with BoundedA at the end) indicates that subjects were randomized uniformly between doing block \textbf{BoundedA} either \textit{before} or \textit{after} all the other blocks in the treatment.  Furthermore, the fact that the boxes containing ``Ellsberg, 2Ball'' and ``EllsbergD, 2BallD'' are adjacent and shaded, in the same way, indicates that, within each of these two randomized groups, there is further randomization as to whether the blocks \textbf{Ellsberg} and \textbf{2Ball} are both completed \textit{before} blocks \textbf{EllsbergD} and \textbf{2BallD} or are both completed after these two blocks.  Finally, in any box containing multiple blocks, those blocks were completed in a random order (e.g., block \textbf{Ellsberg} is either completed before or after block \textbf{2Ball}).  Hence, Figure \ref{fig:structure_nudging} indicates that there are 16 possible orders in which subjects could complete the blocks in treatment \textsc{\bfseries nudging}.

\subsection{Results from Treatment \textsc{\bfseries nudging}}

The first main result of this treatment is that participants prefer more unevenly distributed urns when all involve ambiguity. Table \ref{tab:BoundedA} below summarizes subjects' CEs for the gambles unique to block \textsc{\bfseries nudging}.  Subjects prefer $BB^{60-100}$ to $BB^{40-60}$ by an average of 33.6 cents ($t > 9$). Similarly, they prefer $BB^{95-100}$ to $BB^{60-100}$ by an average of 75.4 cents ($t > 13$).  Of the 179 subjects in treatment \textsc{\bfseries nudging}, only 22 subjects reported a larger CE for $BB^{40-60}$ than $BB^{60-100}$; and even among those 22 subjects, the average CE for $BB^{95-100}$ was massively larger than the average CE for $BB^{60-100}$ (mean of difference = 68.64, $t=3.36$).

\bigskip
\begin{table}[H]
\caption{\sc {\bfseries Bounded-Ambiguity Two-Ball Gambles}}
\centering
\footnotesize
\begin{tabular}{lccc}
\addlinespace
\toprule \toprule
\addlinespace
&\multicolumn{1}{c}{$BB^{40-60}$} & \multicolumn{1}{c}{$BB^{60-100}$}  & \multicolumn{1}{c}{$BB^{95-100}$}  \\
\midrule
Mean &   98.603   &    132.235 &  207.654    \\
SD &   (52.113)          &   (63.887)     & (90.784)                \\
\midrule
$ N $   &     179     &     179    &   179           \\
\addlinespace
\bottomrule \bottomrule
\end{tabular}
\label{tab:BoundedA}
\end{table}
\bigskip

These data indicate that in 2-ball gambles wherein \textit{all} options involve ambiguity, subjects ``correctly'' report significantly larger CEs for gambles drawing from more unevenly distributed urns.  This suggests that the preference for gamble $RR$ over gamble $AA$ may be due to a preference to avoid the mere presence of ambiguity and not due to a lack of understanding that more unevenly distributed urns yield higher win probabilities in 2-ball gambles.

The second main result of this treatment is that participants are not ``nudged'' by bounded ambiguity. One might argue that, while subjects' choices in block \textbf{BoundedA} indicate a significant understanding of the fact that more unevenly distributed urns yield higher win probabilities in 2-ball gambles, this understanding does not exist among subjects who were not exposed to the ``leading questions'' found in block \textbf{BoundedA}.  Indeed, perhaps subjects only come to understand this fact when confronted with stark examples such as an urn containing \textit{at least 95\% red balls}.

Suppose the hypothesis in the previous paragraph was correct. In that case, we should expect to find that the CE difference $R-AA$ is significantly smaller (or, more negative) among subjects who completed block \textbf{BoundedA} \textit{before} completing blocks \textbf{2Ball} and \textbf{2BallD} than it is among subjects who did not complete \textbf{BoundedA} before \textbf{2Ball} and \textbf{2BallD}.  Completing block \textbf{BoundedA} should ``nudge'' subjects into being less susceptible to the 2-ball Ellsberg paradox.

\textit{Half} of the 179 subjects randomly assigned to treatment \textsc{\bfseries nudging} completed \textbf{BoundedA} before the blocks \textbf{2Ball} and \textbf{2BallD}, whereas \textit{none} of the subjects randomly assigned to other treatments did so. So if a nudging effect exists, then it should manifest as a statistically significant (negative) difference between the $R-AA$ values in the treatment \textsc{\bfseries nudging} versus those in the other treatments.

If we let $I^{TN}$ be the indicator variable for assignment to Treatment \textsc{\bfseries nudging}, then in a regression of $Z := R - AA$ on $I^{TN}$, the slope coefficient represents the causal effect of being in Treatment \textsc{\bfseries nudging} on the preference for $R$ over $AA$. A statistically significant \textit{negative} slope coefficient would indicate that Treatment \textsc{\bfseries nudging}  has a nudging effect, causing subjects to manifest less preference for $R$ over $AA$.

\bigskip
\begin{table}[H]
\centering
\footnotesize
\caption{\sc {\bfseries nudging Effects}}
\begin{tabular}{lccc}
\addlinespace
\toprule \toprule
\addlinespace
&\multicolumn{1}{c}{$Z^1$} & \multicolumn{1}{c}{$Z^2$}  & \multicolumn{1}{c}{$Z^{avg}$}  \\
\midrule
$ I^{TN} $ &   2.615   &    -0.418 &  1.098      \\
&   (3.659)          &   (3.917)     & (3.366)                \\
\addlinespace
Const.   &   16.994***   &   16.786***  & 16.890***         \\
&   (1.840)          &   (1.969)     & (1.692)                \\
\midrule
$ N $   &     708     &     708    &   708           \\
\addlinespace
\bottomrule \bottomrule
\end{tabular}
\label{tab:nudging}
\end{table}
\bigskip

Table \ref{tab:nudging} shows the results of such a regression, first using individual elicitations and then the averages across elicitations. As shown, the slope coefficient is not statistically significant, and it is \textit{positive} in the case using averages. Thus, we fail to reject the hypothesis that there is no nudging effect ($p=.63$).  The results of Treatment \textsc{\bfseries nudging} therefore provide strong evidence that subjects' preference for gamble $R$ over gamble $AA$ has little to do with a lack of understanding that gamble $AA$ has a larger win probability.

\section{Is Ambiguity Aversion a Form Complexity Aversion?}\label{sec:complexityaversion}

One might argue that the preference for gamble $R$ over gamble $AA$ is not due to an aversion to the \textit{ambiguity} present in gamble $AA$ but instead to the \textit{complexity} present in gamble $AA$. "Complexity" is a concept difficult to define precisely, and it is not the aim of this paper to do so. However, experiments like \cite{Halevy}'s have established the potential relevance of specific types of complexity, such as the compoundness of lotteries. With this in mind, we test whether the preferences for gamble $R$ over gamble $AA$ is indistinguishable from the preference for a simple 50-50 gamble like $R$ over a \textit{compound} 50-50 gamble, call it $C$ as described in Table \ref{table:gambles_complexity}. We designed Treatment \textsc{\bfseries complexity} to test whether these specific types of complexity may be the primary factors generating the Two-Ball Ellsberg paradox.

The ORIV-corrected correlation between the ``Two-Ball Ellsberg paradox'' preference $R-AA$ and the ``aversion to compound lotteries'' preference $R-C$ is extremely high, but the strength of preference $R-AA$ is greater than that of preference $R-C$.  We find similar results when we compare $R-AA$ to other ``paradoxical'' preferences, such as the classic Ellsberg paradox preference $R-A$.

\subsection{Description of Treatment \textsc{\bfseries complexity}}
The gambles unique to treatment \textsc{\bfseries complexity} are those in block \textbf{Compound}.  In this block, subjects play two gambles involving an urn \textbf{C} that contains 100 balls, all either red or blue.  Subjects are informed that before each gamble begins, the contents of urn \textbf{C} are determined uniformly at random (i.e., each of its 101 possible balls compositions are equally likely to be realized).  Table \ref{table:gambles_complexity} summarizes these gambles.

\bigskip
\begin{table}[H]
\begin{center}
    \begin{tabular}{|c|l|}
    \hline 
        \textsc{gamble name} & \multicolumn{1}{c|}{\textsc{gamble description}} \\
        \cmidrule{1-2}
        \multicolumn{2}{|c|}{Block \textbf{Compound}: Contents of urn \textbf{C} determined uniformly at random} \\
        \hline
        $C$ & Choose a color.  Draw one ball from urn \textbf{C}; win if it's the color you chose. \\
        \hline
         $CC$ & Draw two balls with replacement from urn \textbf{C}; win if they're the same color. \\
        \hline
 \end{tabular}
\caption{\sc {\bfseries description of gambles present only in treatment complexity}}
\label{table:gambles_complexity}
\end{center}
\end{table}
\bigskip

In other words, block \textbf{Compound} consists of two gambles: a compound lottery $C$ and a ``Two-Ball Compound'' gamble $CC$. Gamble $CC$ is the same as the ambiguous gamble $AA$, except its urn's contents are determined by a known lottery rather than an unknown, ambiguous procedure.

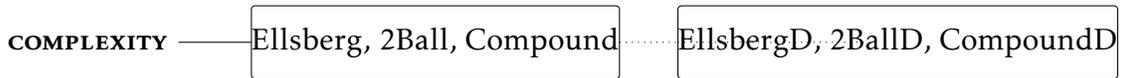
\begin{figure}[H]
\resizebox{15cm}{!}{
\centering
\begin{tikzpicture}
\tikzstyle{block}=[draw, outer sep=0pt, inner sep=0pt, minimum width=5cm, minimum height=1cm, rounded corners=.5mm]
\node[block] (a) {Ellsberg, 2Ball, Compound};
\node[block,right=8mm of a] (b) {EllsbergD, 2BallD, CompoundD};
\coordinate (n) at ([xshift=-1cm]a.west);
 \draw[] (n)--node[left, pos=0]{\textsc{\bfseries complexity}}(a.west);
 \draw[dotted] (a)--(b) (b)--(c);
\end{tikzpicture}}
\caption{\sc {\bfseries structure of treatment complexity}}
\label{fig:structure_complexity}
\end{figure}

Block \textbf{CompoundD} contains duplicate questions of those in block \textbf{Compound}.  In treatment \textsc{\bfseries complexity}, subjects complete the blocks \textbf{Compound}, \textbf{Ellsberg} and \textbf{2Ball} as well as the duplicate blocks \textbf{CompoundD}, \textbf{EllsbergD} and \textbf{2BallD}.  The order in which these blocks were presented was determined at random, independently for each subject assigned to this treatment, according to Figure \ref{fig:structure_complexity}.  Its interpretation is analogous to that of Figure \ref{fig:structure_nudging}; there are 12 different orders in which the six blocks comprising Treatment \textsc{\bfseries complexity} could be completed. 
Table \ref{tab:rawvariables} in Appendix \ref{sec:appendixraw} contains summary statistics for each elicitation of CEs for gambles $C$ and $CC$.

\bigskip

\subsection{Results from Treatment \textsc{\bfseries complexity}}

The variable $R-A$ measures subjects' ambiguity aversion in the classic Ellsberg paradox, while $R-RR$ measures their preference for a simple 50-50 gamble to a Two-Ball 50-50 gamble. $R-C$ measures subjects' preference for a simple 50-50 gamble over a Compound 50-50 gamble, and $R-CC$ measures their preference for a simple 50-50 gamble over a Two-Ball Compound 50-50 gamble.  Table \ref{tab:derivedvariables} in Appendix \ref{sec:appendixderived} contains summary statistics for each elicitation of these CE differences.

Table \ref{orivcorrzehl} computes the ORIV-adjusted correlations between our central variable $R-AA$ and these other variables.

\begin{table}[H]
\centering
\footnotesize
\caption{\sc {\bfseries relationships between ce differences}}
\begin{tabular}{lcccccccc}
\addlinespace
\toprule \toprule
\addlinespace
\multicolumn{9}{c}{Dependent Variable: $ R-AA $} \\
\midrule

Indep. Variable: & & $ R-A $  & &  $ R-RR $    & & $ R-C $  & & $ R-CC $ \\
\midrule
ORIV $ \rho $  & & \multicolumn{1}{l}{0.892        }  & & \multicolumn{1}{l}{0.952         } & & \multicolumn{1}{l}{0.954 } & & \multicolumn{1}{l}{0.917 } \\
& & \multicolumn{1}{l}{(0.017)     }  & & \multicolumn{1}{l}{(0.012)  } & & \multicolumn{1}{l}{(0.024) } & & \multicolumn{1}{l}{(0.032) }\\
\addlinespace
\Xhline{1\arrayrulewidth}
$ N $      & & 708        & & 708         & & 158 & & 158 \\
\addlinespace
\bottomrule \bottomrule
\end{tabular}
\label{orivcorrzehl}
\end{table}

As the table shows, the preference for $R$ over $AA$ is extremely tightly correlated with each of the preferences mentioned in the previous paragraph.  Thus, from the analyst's point of view, a subject exhibiting one of these ``paradoxical'' preferences to a certain degree of strength (as measured by standard deviations above the population mean) makes it exceedingly likely that she will exhibit these other ``paradoxical'' preferences to a similar degree of strength. In particular, this finding replicates \cite{Halevy}'s and \cite{GSY}'s conclusions that ambiguity aversion in the classic Ellsberg paradox is tightly linked to failure to reduce compound lotteries.

Besides correlations, it is worthwhile to examine the \textit{differences} between the variables in the table above.  $R-AA$ is larger than all of $R-A$, $R-RR$, and $R-C$ ($t > 4$ in all cases) and is larger than $R-CC$ by a statistically insignificant amount ($t = 1.05$).  This suggests that, according to most subjects, gamble $AA$ is likely the ``worst'' of gambles $AA$, $A$, $RR$, $C$, and $CC$ - perhaps because gamble $AA$ combines ambiguity and Two-Ball complexity.  The only possible competitor for being the ``worst'' is gamble $CC$, which is identical to gamble $AA$ except that its urn's contents are determined \textit{randomly} rather than in an ambiguous manner.

\section{Do Subjects Believe Urn Contents Can Change Between Draws?}\label{sec:robustness}
Another hypothesis that might explain a preference for gamble $R$ over gamble $AA$ is that subjects incorrectly believe that the contents of urn \textbf{A} can change between the two draws (with replacement) made from it in gamble $AA$.  Let us call this the \textit{False Independence} hypothesis.  If this hypothesis were true, then gamble $AA$ need not have a higher win probability than gamble $R$, and may be less desirable since its win probability is ambiguous while that of gamble $R$ is not.  We designed Treatment \textsc{\bfseries robustness} to test the False Independence hypothesis, as well as to contain exploratory gambles meant to motivate further research that will be discussed in Section \ref{sec:distaste}.\footnote{Due to constraints on the maximum number of gambles we could fit in a given treatment, we could not fit all such exploratory gambles in a single treatment.  Thus, we split them across two treatments.}

\subsection{Description of Treatment \textsc{\bfseries robustness}}\label{sec:treatmentrobustness}
The gambles unique to treatment \textsc{\bfseries robustness} are those in blocks \textbf{Independent} and \textbf{3Ball}.  In block \textbf{Independent}, subjects draw a ball from each of two ambiguous urns (containing only red and blue balls) whose contents were determined independently; they win \$3 if the two balls have the same color.  In block \textbf{3Ball}, subjects draw 3 balls in total, with replacement, from some combination of a single ambiguous urn \textbf{A} and a single risky urn \textbf{R}, in a certain order.  They win \$3 if \textit{all three} balls have the same color.  The gambles in these blocks are summarized in Table \ref{table:gambles_robustness}.

\bigskip
\begin{table}[H]
\begin{center}
    \begin{tabular}{|c|l|}
    \hline 
        \textsc{gamble name} & \multicolumn{1}{c|}{\textsc{gamble description}} \\
        \cmidrule{1-2}
        \multicolumn{2}{|c|}{Block \textbf{Independent}: Win if two balls drawn from separate urns have same color} \\
        \hline
        $IA$ & Two ambiguous urns containing only red and blue balls; \\
        & their contents were determined independently. \\
        \hline
        \multicolumn{2}{|c|}{Block \textbf{3Ball}: Win if three balls drawn with replacement have same color} \\
        \hline
        $RRR$ & 1st ball from urn \textbf{R}; 2nd ball from urn \textbf{R}; 3rd ball from urn \textbf{R}. \\
        \hline
         $AAA$ & 1st ball from urn \textbf{A}; 2nd ball from urn \textbf{A}; 3rd ball from urn \textbf{A}. \\
        \hline
         $RAA$ & 1st ball from urn \textbf{R}; 2nd ball from urn \textbf{A}; 3rd ball from urn \textbf{A}. \\
         \hline
 \end{tabular}
\caption{\sc {\bfseries description of gambles present only in treatment robustness}}
\label{table:gambles_robustness}
\end{center}
\end{table}
\bigskip

\begin{figure}[H]
\resizebox{18cm}{!}{
\begin{tikzpicture}
\tikzstyle{block}=[draw, outer sep=0pt, inner sep=0pt, minimum width=5cm, minimum height=1cm, rounded corners=.5mm]
\node[block] (a) {Ellsberg, 2Ball};
\node[block,right=8mm of a] (b) {EllsbergD, 2BallD};
\node[block,right=8mm of b] (c) {3Ball, Independant};
\coordinate (n) at ([xshift=-1cm]a.west);
 \draw[] (n)--node[left, pos=0]{\textsc{\bfseries robustness}}(a.west);
 \draw[dotted] (a)--(b) (b)--(c);
\end{tikzpicture}}
\caption{\sc {\bfseries structure of treatment robustness}}
\label{fig:structure_robustness}
\end{figure}
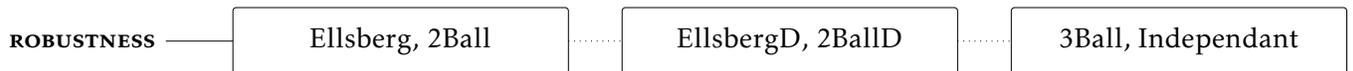

In treatment \textsc{\bfseries robustness}, subjects complete the blocks \textbf{Independent}, \textbf{3Ball}, \textbf{Ellsberg} and \textbf{2Ball} as well as the duplicate blocks \textbf{EllsbergD} and \textbf{2BallD}.  The order in which these blocks were presented was determined at random, independently for each subject assigned to this treatment, according to Figure \ref{fig:structure_robustness}.  Its interpretation is analogous to that of Figure \ref{fig:structure_nudging}; there are 48 different orders in which the six blocks comprising Treatment \textsc{\bfseries robustness} could be completed.
\bigskip
\bigskip
In gamble $IA$, the only one in block \textbf{Independent}, the mean of the subjects' CEs was 107.839, and its standard deviation was 68.733.  We discuss the gambles in block \textbf{3Ball} in Section \ref{sec:distaste}.

\subsection{Results from Block \textbf{Independent}}\label{sec:IndepResults}

On average, the 192 subjects in Treatment \textsc{\bfseries robustness} slightly preferred $AA$ to $IA$, but the difference is not statistically significant (mean = 1.73, $t=.60$).  An indifference between $AA$ and $IA$ would be consistent with the False Independence hypothesis, while a strict preference for $AA$ over $IA$ would falsify this hypothesis.  We, therefore, fail to reject the False Independence hypothesis. A future experiment that replicates block \textbf{Independent} with a larger sample size or larger payments may be able to falsify it; or, see Section \ref{sec:discussion} for discussion of a variation on block \textbf{BoundedA} that may be able to shed further light on this hypothesis.

\section{Does the Mere Presence of Ambiguity, or the Amount of Ambiguity, Matter?}\label{sec:distaste}
A further hypothesis that could explain a preference for gamble $R$ over gamble $AA$ is that the \textit{mere presence} of ambiguity in gamble $AA$ makes it undesirable to subjects or that, more generally, gambles containing a larger ``amount'' of ambiguity are less preferable (all else equal).  The purpose of this paper is not to precisely articulate what a ``distaste for the mere presence of ambiguity'' means or to define the ``amount of ambiguity'' present in a gamble, and it is difficult to separate concepts like these from a distaste for the presence or amount of complexity.  

Hence, it is difficult to test the hypothesis that the Two-Ball Ellsberg paradox is due to a distaste for the \textit{amount of ambiguity present in gamble} $AA$. Nonetheless, we expected that eliciting subjects' CEs for certain gambles closely related to $AA$ - namely gambles $AR$ and $RA$ described below and gambles $AAA$, $RRR$ and $RAA$ from Section \ref{sec:treatmentrobustness} - may shed light on this hypothesis.  We thus designed Treatment \textsc{\bfseries order}, and block \textbf{3Ball} from Treatment \textsc{\bfseries robustness}, to provide additional information that may motivate future research.

\subsection{Description of Treatment \textsc{\bfseries order}}\label{sec:level}

The gambles unique to treatment \textsc{\bfseries order} are found in block \textbf{2BallMixed} - an expanded version of block \textbf{2Ball} that contains gambles not only $RR$ and $AA$ as before but also gambles $AR$ and $RA$.  In each gamble, subjects draw two balls - either from the same urn and with replacement or from distinct urns - in a certain order, and they win \$3 if the two balls have the same color.  Table \ref{table:gambles_level} summarizes these gambles.

\bigskip
\begin{table}[H]
\begin{center}
    \begin{tabular}{|c|l|}
    \hline 
        \textsc{gamble name} & \multicolumn{1}{c|}{\textsc{gamble description}} \\
        \cmidrule{1-2}
        \multicolumn{2}{|c|}{Block \textbf{2BallMixed}: Win if two balls drawn with replacement have same color} \\
        \hline
        $RR$ & 1st ball from urn \textbf{R}; 2nd ball from urn \textbf{R}. \\
        \hline
        $AA$ & 1st ball from urn \textbf{A}; 2nd ball from urn \textbf{A}. \\
        \hline
        $AR$ & 1st ball from urn \textbf{A}; 2nd ball from urn \textbf{R}. \\
        \hline
        $RA$ & 1st ball from urn \textbf{R}; 2nd ball from urn \textbf{A}. \\
        \hline
 \end{tabular}
\caption{\sc {\bfseries description of gambles present in treatment order}}
\label{table:gambles_level}
\end{center}
\end{table}
\bigskip

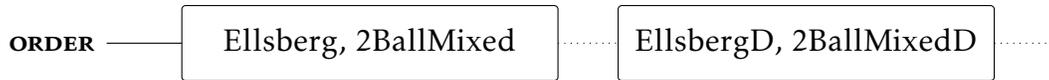
\begin{figure}[H]
\begin{tikzpicture}
\tikzstyle{block}=[draw, outer sep=0pt, inner sep=0pt, minimum width=5cm, minimum height=1cm, rounded corners=.5mm]
\node[block] (a) {Ellsberg, 2BallMixed};
\node[block,right=8mm of a] (b) {EllsbergD, 2BallMixedD};
\coordinate (n) at ([xshift=-1cm]a.west);
 \draw[] (n)--node[left, pos=0]{\textsc{\bfseries order}}(a.west);
 \draw[dotted] (a)--(b) (b)--(c);
\end{tikzpicture}
\caption{\sc {\bfseries structure of treatment order}}
\label{fig:structure_level}
\end{figure}

Block \textbf{2BallMixedD} contains duplicate questions of those in block \textbf{2BallMixed}.  In treatment \textsc{\bfseries order}, subjects complete the blocks \textbf{2BallMixed} and \textbf{Ellsberg} as well as the duplicate blocks \textbf{2BallMixedD} and \textbf{EllsbergD}.  The order in which these blocks were presented was determined at random, independently for each subject assigned to this treatment, according to Figure \ref{fig:structure_level}.  Its interpretation is analogous to that of Figure \ref{fig:structure_nudging}; there are 8 different orders in which the 4 blocks comprising Treatment \textsc{\bfseries order} could be completed.

\bigskip
\bigskip

\subsection{Results From Treatment \textsc{\bfseries order}}\label{sec:paradoxesresults}
Table \ref{tab:rawvariables} in Appendix \ref{sec:appendixraw} contains summary statistics for each elicitation of CEs for gambles $RR$, $AA$, $AR$, and $RA$. Besides, table \ref{tab:derivedvariables} in Appendix \ref{sec:appendixderived} contains summary statistics for each elicitation of the differences between the CEs $RR$, $AA$, $AR$, and $RA$.  

The average CEs for the four gambles in block \textbf{2BallMixed} are ranked in the order
\[ RR > AA > AR > RA, \]
but the only statistically significant differences between these variables are those between $RR$ and each of the other three.  Hence, we cannot rule out the possibility that subjects' average preferences are of the form
\[ RR \succ AA \sim AR \sim RA. \]

It is worth noting that gambles $RA$ and $AR$ each have a win probability of exactly 50\%: whatever ball is drawn from urn \textbf{A}, the ball from urn \textbf{R} has a 50\% chance of matching it.  Hence, if subjects care only about win probabilities and exhibit no aversion to the mere presence of ambiguity, then we should expect a preference ordering like $AA \succsim RR \sim RA \sim AR$.  Meanwhile, if subjects do not care at all about win probabilities and respond solely based on a distaste for the mere presence of ambiguity, then we should expect a preference ordering like $RR \succ AR \sim RA \succ AA.$\footnote{Here, we have informally formulated a ``distaste for the mere presence of ambiguity'' to be such that each additional instance of a draw from an ambiguous urn makes the overall gamble more distasteful.} 
A strict preference for RR over RA may also be explainable by subjects failing to realize that the preference of urn R automatically hedges against the ambiguity in urn A and causes gamble RA to have an unambiguous probability of winning of 50\%. Lastly, strict preference between $AR$ and $RA$ would indicate \textit{order sensitivity} and that a more intricate explanation is required.

Although we do not find significant evidence of order sensitivity, we find that neither a total indifference to the mere presence of ambiguity nor a preference based solely on avoiding the mere presence of ambiguity, is sufficient to explain subjects' behavior.  Indeed, gamble $AA$ is neither at least as good as all three other gambles (as the first simplistic theory would predict) nor worse than all three other gambles (as the second would predict).  We speculate that subjects may find gamble $AA$ to be at least as good as gambles $AR$ and $RA$ because its larger probability of winning (than the 50\% given by $AR$ or $RA$) offsets its increased presence of ambiguity, but find $AA$ to be worse than $RR$ because $RR$'s complete lack of ambiguity makes it significantly more attractive.

\subsection{Results from Block \textbf{3Ball}}
Recall the gambles in block \textbf{3Ball} summarized in Table \ref{table:gambles_robustness} above.  Table \ref{tab:3Ball} presents summary statistics of subjects' CEs for these gambles.

\begin{table}[H]
\caption{\sc {\bfseries CEs for 3-Ball gambles}}
\centering
\footnotesize
\begin{tabular}{lccc}
\addlinespace
\toprule \toprule
\addlinespace
&\multicolumn{1}{c}{$RRR$} & \multicolumn{1}{c}{$AAA$}  & \multicolumn{1}{c}{$RAA$}  \\
\midrule
Mean &   97.708   &    91.120 &  92.552      \\
SD &   (67.310)          &   (69.264)     & (68.172)                \\
\midrule
$ N $   &     192     &     192    &   192           \\
\addlinespace
\bottomrule \bottomrule
\end{tabular}
\label{tab:3Ball}
\end{table}

These reported CEs are too large for a classical risk-averse agent who correctly calculates the probabilities of winning.\footnote{Our results from the simple 50-50 gamble in block \textbf{Ellsberg} suggest that subjects are on average slightly risk averse.} Notice that $RRR$ has a win probability of exactly $\frac{1}{4}$, but subjects report an average CE of 97.7 cents for this gamble - a value significantly larger than the risk-neutral CE of 75 cents ($t=4.67$).  Similarly, subjects on average value gamble $RAA$ significantly at more than half as much as gamble $AA$ (difference of means = 37.77, $t=11.03$). Thus, subjects seemingly \textit{overweight} the win probabilities of 3-Ball gambles.

Despite the general overweighting of win probabilities, comparisons between CEs for these 3-Ball gambles remain qualitatively similar to the comparisons between the CEs for Two-Ball gambles. Similarly to how subjects on average preferred $RR$ to $AA$, we find that subjects on average prefer $RRR$ to $AAA$ (mean = 6.59, $t=2.16$), even though $AAA$ must have \textit{at least as large} of a win probability as $RRR$.  Also, just as we did not find a statistically significant difference between $RA$ and $AA$ in Section \ref{sec:paradoxesresults}, we find no statistically significant difference between $RAA$ and $AAA$ (mean = 1.43, $t=.56$). 

Likewise, these results suggest that perhaps the ``amount'' of ambiguity present in a gamble, measured in terms of the total number of (or proportion of) draws that come from ambiguous urns, does not matter as much as the \textit{mere presence of ambiguity at all}.

\section{Discussion}\label{sec:discussion}

\paragraph{Distaste for Ambiguity or Complexity?}
Our treatment \textsc{\bfseries nudging} suggests that individuals' choice of gamble $RR$ over $AA$ is not due entirely to a lack of cognition or a failure to reduce compound lotteries.  Nonetheless, one might argue that the preference for $RR$ over $AA$ is due to a distaste for \textit{complexity} rather than ambiguity.  Even in this case, we have identified the \textit{mere presence of ambiguity} as a driver of change in people's behavior, perhaps through the complexity, it introduces or perhaps through other means.\footnote{If ambiguity is simply a type of complexity, then models of contingent reasoning such as \cite{martinez2019failures} may be good candidates to explain our results.}

One might further suggest that the preference for $RR$ over $AA$ comes from ''inappropriate'' beliefs, i.e., beliefs that are not the product of the same distribution over the contents of the urn.  Our treatment \textsc{\bfseries robustness} was designed to address this concern but its results lacked the statistical significance to rule out this explanation.

Whether explained or not as an instance of complexity, people harboring a distaste for the mere presence of ambiguity has potentially widespread implications for economics. Subjects may prefer to gamble $R$ to $A$ in the classic Ellsberg paradox primarily because they dislike the mere presence of ambiguity and not, for instance, entirely because they hold concern for worst-case scenarios, as \cite{GilboaSchmeidler} would suggest. Models ignoring a distaste for ambiguity \textit{per se} would incorrectly predict individuals' behavior in a variety of situations.  Hence, new models may be required.

\paragraph{Raiffa Critique.}
Unlike in the original Ellsberg paradox, a subject cannot eliminate the ambiguity present in gamble $AA$ by introducing randomization in her choice of color (as in \cite{raiffa1961risk}). Indeed, gamble $AA$ does not ask subjects to choose a color. Even if we presented subjects with a modified version of gamble $AA$ wherein they choose either red or blue and win if and only if both balls drawn were of the chosen color (and compared this to a similarly modified version of gamble $RR$), it is still the case that randomizing one's color choice does not eliminate the ambiguity in the payoff of gamble $AA$. If $p$ is the (ambiguous) proportion of red balls in urn \textbf{A}, then this modified version of gamble $AA$ has win probability $p^2$ when you bet on red and win probability $(1-p)^2$ when you bet on blue. 

Randomizing your choice of color 50-50 would thus mean that the gamble's win probability is $.5p^2 + .5(1-p)^2 \geq .25$. In contrast, the modified version of gamble $RR$ has a probability of winning $.25$ regardless of the color on which you bet (or whether you randomized your choice of color). It is still the case that gamble $AA$ has an ambiguous win probability and that it is at least as large as (and in all but one case, strictly larger than) that of $RR$.

\paragraph{A different experiment to reject the False Independence hypothesis.} Recall the "False Independence" hypothesis mentioned in Section \ref{sec:robustness}: Do subjects imagine that our "two draws with replacement from the same ambiguous urn" are actually "two draws from two ambiguous urns whose contents were determined independently"?  Our experiment can't  rule out the False Independence assumption as a driving factor in the 2-Ball Ellsberg paradox, but here we suggest how a further experiment might do so.

A variation on block \textbf{BoundedA} may be sufficient to show that the False Independence assumption cannot fully explain our results. Consider a version of gamble $BB^{95-100}$ wherein instead of the gamble specifying that the urn contains between 95 and 100 \textit{red} balls, it merely specifies that \textit{at least 95 of the 100 balls in the urn are of the same color}. Suppose subjects imagined the two draws from the specified urn as ``one draw from each of two distinct urns, whose contents were determined in a specified manner but were determined independently.'' Then we should not find a strong preference for this version of gamble $BB^{95-100}$ over gamble $AA$.  

Indeed, suppose subjects believe in False Independence. In that case, they might easily imagine this new version of gamble $BB^{95-100}$ to have a win probability close to 50\%. For although it is possible in their minds that "both urns" contain at least 95 red balls (or that both contain at least 95 blue balls), it is equally possible to them that "one urn contains at least 95 red balls while the other contains at least 95 blue balls". In other words, their CEs for this version of gamble $BB^{95-100}$ should certainly \textit{not} be radically larger than their CEs for gamble $AA$.  If such a radical difference in CEs as we found between the original version of gamble $BB^{95-100}$ and gamble $AA$ were still found under this modified version of $BB^{95-100}$, this would suggest that False Independence is not the primary factor generating our results.

\section{Conclusion}\label{sec:conclusion}

Two-Ball gambles are a rich class of decision problems. Because they can involve ambiguity but guarantee a minimum win probability that is at least as large as that of some other gamble, they allow us to test whether subjects avoid ambiguity \textit{per se} as opposed to avoiding ambiguity because it may yield a worse outcome.

The most striking case of preferring a gamble with lower win probability is that subjects preferred the 50-50 gamble, $R$, to the Two-Ball ambiguous gamble, $AA$. This preference is closely correlated with the traditional Ellsberg preference for $R$ over a 1-Ball ambiguous gamble $A$, and also with the preference for $R$ over the compound 50-50 gamble $C$, as well as the preference for $R$ over the Two-Ball 50-50 gamble $RR$. These close relationships suggest that it may be difficult to separate an aversion to ambiguity \textit{per se} from an aversion to complexity.

It is implausible that subjects prefer $R$ to $AA$ simply due to a poor understanding of Two-Ball gambles. In the block \textbf{BoundedA}, subjects correctly and strongly identified that more unevenly distributed urns are more likely to win. Moreover, the lack of a "nudging" effect from being in the treatment containing \textbf{BoundedA} suggests that subjects' preference for $R$ over $AA$ is \textit{deliberate}.

Although the presence of an ambiguous draw within a gamble is associated with a significantly lower CE, it remains unclear whether having more ambiguity, as measured perhaps by the \textit{number} or \textit{proportion} of ambiguous draws present in a gamble, has an additional negative effect on the CE. This presents an interesting question for further research.

\singlespacing
\bibliography{2Ellsberg_Bibliography.bib}

\newpage
\appendix

\begin{center}
\section*{Appendix}
\end{center}

\section{Main Tables}

\subsection{Raw Variable Names}
\begin{table}[H]
\centering
\caption{\sc {\bfseries raw variable names}}
\begin{adjustbox}{max width=\textwidth}
\small
  \begin{tabular}{@{}ll@{}}
   \toprule \toprule
\textbf{Name} & \textbf{Description} \\
\midrule
\addlinespace
$R^j$ & $j$th elicitation of CE for 50-50 urn of \textbf{Ellsberg}\\
$A^j$ & $j$th elicitation of CE for ambiguous urn of \textbf{Ellsberg}\\
\addlinespace
\Xhline{1\arrayrulewidth}
\addlinespace
$RR^j$ & $j$th elicitation of CE for 50-50 urn in \textbf{2BallMixed}\\
$AA^j$ & $j$th elicitation of CE for ambiguous urn in \textbf{2BallMixed}\\ 
$AR^j$ & $j$th elicitation of CE for "1st urn=\textbf{A}, 2nd=\textbf{R}" gamble of \textbf{2BallMixed} \\ 
$RA^j$ & $j$th elicitation of CE for "1st urn=\textbf{R}, 2nd=\textbf{A}" gamble of \textbf{2BallMixed} \\
\addlinespace
\Xhline{1\arrayrulewidth}
\addlinespace
$R3$ & CE for \textbf{3Ball} with all three urns = \textbf{R} \\
$A3$ & CE for \textbf{3Ball} with all three urns = \textbf{A} \\
$RAA$ & CE for \textbf{3Ball} with 1st urn = \textbf{R}, latter two urns = \textbf{A}   \\
\addlinespace
\Xhline{1\arrayrulewidth}
\addlinespace
$IA$ & CE for \textbf{Independent} (Two-Ball gamble with independent ambiguous urns) \\
\addlinespace
\Xhline{1\arrayrulewidth}
\addlinespace
$C^j$ & $j$th elicitation of CE for single-urn gamble of \textbf{Compound} \\
$CC^j$& $j$th elicitation of CE for Two-Ball gamble of \textbf{Compound} \\
\addlinespace
\Xhline{1\arrayrulewidth}
\addlinespace
$BB^{40-60}$ & CE for \textbf{BoundedA} with ambiguous urn containing 40-60 red balls\\
$BB^{60-100}$ & CE for \textbf{BoundedA} with ambiguous urn containing 60-100 red balls  \\
$BB^{95-100}$ & CE for \textbf{BoundedA} with ambiguous urn containing 95-100 red balls \\[2pt] 
\bottomrule \bottomrule
  \end{tabular}
  \end{adjustbox}
  \label{tab:mainvarnames}
\end{table}

\subsection{Summary Statistics for Raw Variables}\label{sec:appendixraw}
\begin{sidewaystable}[H]
\centering
\footnotesize
\caption{\sc {\bfseries raw variables: decomposed summary statistics}}
\resizebox{22cm}{!}{
\begin{tabular}{lcccccccccccc}
\addlinespace
\toprule \toprule
\addlinespace
& &     \multicolumn{2}{c}{$R$} & & \multicolumn{2}{c}{$A$} && \multicolumn{2}{c}{$C$} && \multicolumn{2}{c}{$RA$} \\
\cmidrule{1-1}           \cmidrule{3-4} \cmidrule{6-7} \cmidrule{9-10} \cmidrule{12-13}
Mean                    & &       118.55    &     117.72     &&    105.42      &    106.22     & &     108.67     &     111.55    & &    95.67    &     93.72     \\ 
$ 95\% $ Conf. Interval & &   [114.43 , 122.66 ] &  [113.48 , 121.96 ] &&  [101.10 , 109.74 ] & [101.81 , 110.63 ] & &  [100.15 , 117.19 ] &  [102.31 , 120.79 ] & & [ 87.64 , 103.70 ] & [ 85.54 , 101.89 ] \\
\cmidrule{1-1} \cmidrule{3-4} \cmidrule{6-7} \cmidrule{9-10} \cmidrule{12-13}
$ \rho $              & &  \multicolumn{2}{c}{ 0.875 }      & &  \multicolumn{2}{c}{ 0.871 }      & &     \multicolumn{2}{c}{ 0.878 }       & &   \multicolumn{2}{c}{ 0.855 }       \\
& &  \multicolumn{2}{c}{ (0.018) }      & &  \multicolumn{2}{c}{ (0.018) }      & &     \multicolumn{2}{c}{ (0.038) }        & &   \multicolumn{2}{c}{ (0.039) }       \\
\addlinespace
\Xhline{0.1\arrayrulewidth}
$ N $            & &  \multicolumn{2}{c}{ 708 }       & &  \multicolumn{2}{c}{ 708 }       & &     \multicolumn{2}{c}{ 158 }        & &   \multicolumn{2}{c}{ 179 }              \\
\addlinespace
\Xhline{2.5\arrayrulewidth}
\addlinespace
& &     \multicolumn{2}{c}{$RR$} & & \multicolumn{2}{c}{$AA$} && \multicolumn{2}{c}{$CC$} && \multicolumn{2}{c}{$AR$} \\
\cmidrule{1-1}           \cmidrule{3-4} \cmidrule{6-7} \cmidrule{9-10} \cmidrule{12-13}
Mean                    & &       110.55    &     108.64     &&    100.95      &    101.12     & &       102.53     &     104.97    & &       97.21    &     95.78     \\
$ 95\% $ Conf. Interval & &   [106.25 , 114.85 ] &  [104.44 , 112.85 ] &&  [ 96.59 , 105.30 ] & [ 96.60 , 105.64 ] & &  [ 93.51 , 111.56 ] &  [ 95.49 , 114.45 ] & & [ 89.15 , 105.26 ] & [ 87.61 , 103.95 ] \\
\cmidrule{1-1} \cmidrule{3-4} \cmidrule{6-7} \cmidrule{9-10} \cmidrule{12-13}
$ \rho $              & &  \multicolumn{2}{c}{ 0.866 }      & &  \multicolumn{2}{c}{ 0.876 }      & &     \multicolumn{2}{c}{ 0.855 }       & &   \multicolumn{2}{c}{ 0.883 }       \\
& &  \multicolumn{2}{c}{ (0.019) }      & &  \multicolumn{2}{c}{ (0.018) }      & &     \multicolumn{2}{c}{ (0.042) }        & &   \multicolumn{2}{c}{ (0.035) }       \\
\addlinespace
\Xhline{0.1\arrayrulewidth}
$ N $            & &  \multicolumn{2}{c}{ 708 }       & &  \multicolumn{2}{c}{ 708 }       & &     \multicolumn{2}{c}{ 158 }        & &   \multicolumn{2}{c}{ 179 }       \\
\addlinespace
\bottomrule \bottomrule
\end{tabular}}
\label{tab:rawvariables}
\end{sidewaystable}

\subsection{Summary Statistics for Derived Variables}\label{sec:appendixderived}
\begin{sidewaystable}[H]
\centering
\small
\caption{\sc {\bfseries derived variables: decomposed summary statistics}}
\resizebox{22cm}{!}{
\begin{tabular}{lcccccccccccc}
\addlinespace
\toprule \toprule
\addlinespace
& &     \multicolumn{2}{c}{$R-A$} & & \multicolumn{2}{c}{$R-C$} && \multicolumn{2}{c}{$R-AA$} && \multicolumn{2}{c}{$R-CC$} \\
\cmidrule{1-1}           \cmidrule{3-4} \cmidrule{6-7} \cmidrule{9-10} \cmidrule{12-13}
Mean                    & &        13.15    &     11.54     &&     7.91      &     5.70     & &      17.66     &     16.68    & &    14.11    &      12.34     \\
$ 95\% $ Conf. Interval & &   [ 10.46 , ] &  [ 8.64 , 14.44 ] &&  [ 3.37 , 12.45 ] & [ 0.23 , 11.16 ] & &  [ 14.53 , 20.78 ] &  [ 13.34 , 20.02 ] & & [ 6.91 , 21.32 ] & [ , 19.82 ] \\
\cmidrule{1-1} \cmidrule{3-4} \cmidrule{6-7} \cmidrule{9-10} \cmidrule{12-13}
$ \rho $              & &  \multicolumn{2}{c}{ 0.489 }      & &  \multicolumn{2}{c}{ 0.364 }      & &     \multicolumn{2}{c}{ 0.578 }       & &   \multicolumn{2}{c}{ 0.561 }       \\
& &  \multicolumn{2}{c}{ (0.033) }      & &  \multicolumn{2}{c}{ (0.075) }      & &     \multicolumn{2}{c}{ (0.031) }        & &   \multicolumn{2}{c}{ (0.066) }       \\
\addlinespace
\Xhline{0.1\arrayrulewidth}
$ N $            & &  \multicolumn{2}{c}{ 708 }       & &  \multicolumn{2}{c}{ 158 }       & &     \multicolumn{2}{c}{ 708 }        & &   \multicolumn{2}{c}{ 158 }       \\
\addlinespace
\Xhline{2.5\arrayrulewidth}
\addlinespace
& &     \multicolumn{2}{c}{$R-RR$} & & \multicolumn{2}{c}{$RA-AR$} && \multicolumn{2}{c}{$RA-AA$} && \multicolumn{2}{c}{$RA-RR$} \\
\cmidrule{1-1}           \cmidrule{3-4} \cmidrule{6-7} \cmidrule{9-10} \cmidrule{12-13}
Mean                    & &        8.02    &      9.08     &&     -1.51      &    -2.07     & &      0.84     &     -2.85    & &    -11.68    &     -11.01     \\
$ 95\% $ Conf. Interval & &   [ 5.33 , ] &  [ 6.40 , 11.77 ] &&  [ -4.85 ,  1.84 ] & [ -5.48 ,  1.35 ] & &  [ -4.76 ,  6.44 ] &  [ -7.58 ,  1.87 ] & & [-16.81 , -6.55 ] & [-15.47 , -6.54 ] \\
\cmidrule{1-1} \cmidrule{3-4} \cmidrule{6-7} \cmidrule{9-10} \cmidrule{12-13}
$ \rho $              & &  \multicolumn{2}{c}{ 0.428 }      & &  \multicolumn{2}{c}{ 0.151 }      & &     \multicolumn{2}{c}{ 0.368 }       & &   \multicolumn{2}{c}{ 0.259 }       \\
& &  \multicolumn{2}{c}{ (0.034) }      & &  \multicolumn{2}{c}{ (0.074) }      & &     \multicolumn{2}{c}{ (0.070) }        & &   \multicolumn{2}{c}{ (0.073) }       \\
\addlinespace
\Xhline{0.1\arrayrulewidth}
$ N $            & &  \multicolumn{2}{c}{ 708 }       & &  \multicolumn{2}{c}{ 179 }       & &     \multicolumn{2}{c}{ 179 }        & &   \multicolumn{2}{c}{ 179 }       \\

\addlinespace
\bottomrule \bottomrule
\end{tabular}}
\label{tab:derivedvariables}
\end{sidewaystable}

\section{Future Research: Is it really a distaste for the mere presence of ambiguity?}
\subsection{New Experiment 1}

\paragraph{Description of Gamble.}"Urn \textbf{R} has 50 red and 50 blue; urn \textbf{A} has 100 balls in total, all red or blue, with \textit{at least 50 of them red}.  You win if you draw a red ball.  Do you prefer to play this gamble with urn \textbf{R} or urn \textbf{A}?"  Or, perhaps we guarantee instead that "Urn \textbf{A} contains between 50 and 60 red balls; the rest of its 100 balls are blue."

\paragraph{Expected finding.} People prefer urn \textbf{A} since it has at least as high of a chance of winning as urn \textbf{R} does.

\paragraph{Possible Critique from this finding.} People don't exhibit any distaste for the mere presence of ambiguity; they merely fail to calculate odds correctly when you make things opaque/complicated enough.  All of our 2Ellsberg findings are an artifact of the fact that we've framed the gambles one way rather than a more straightforward way.

\paragraph{Responses to these critiques.}

Notice that people \textit{do} "correctly" identify that $BB^{95-100} \succ BB^{60-100} \succ BB^{40-60}$.  Furthermore, their preference for $RR$ over $AA$ is robust to being "nudged" by the \textbf{BoundedA} block.  This all suggests that the original preference for $RR$ over $AA$ cannot entirely be due to "a lack of understanding that more unequal urns are better in a 2-ball gamble."  

But what, then, could explain why our results show a distaste for 'ambiguity that can only help you' while New Experiment 1 shows the opposite?  Perhaps the key difference is that New Experiment 1 frames things in a way that immediately suggests a probabilistic dominance of urn \textbf{A} over urn \textbf{R}, while our $AA$ vs. $RR$ question does not.  Indeed, perhaps most people do not employ probabilistic thinking in pretty much any scenarios - they only use probabilities when "forced" to do so by the odds of winning being given to them (nearly) explicitly.  A comparison between urns \textbf{A} and \textbf{R} in New Experiment 1 forces the observation that "the minimum win probability in urn \textbf{A} is at least as high as the win probability in urn \textbf{R}," but in 2Ellsberg it does not suggest this observation since the conditional win probabilities (for each ball composition of urn \textbf{A}) are 'hidden'.

\subsection{New Experiment 2}

\paragraph{Description of Gamble.} Elicit CEs for an $AA$ gamble but this time specify that urn $A$ has one of the following three ball compositions:
\begin{itemize}
    \item 50 red balls and 50 blue balls.
    \item 75 red balls and 25 blue balls.
    \item 25 blue balls and 75 red balls.
\end{itemize}
Also, elicit people's CEs for 2-ball gambles from risky urns (call them urns \textbf{R}, \textbf{S}, and \textbf{T}) that are 50-50, 75-25 and 25-75 in composition.  Randomize the order of whether you ask about gambles $RR$, $SS$ and $TT$ before or after gambles $AA$ and $RR$.

In each of the 75-25 cases, urn \textbf{A} has a .625 probability of winning.  It would be interesting (and a counterexample to Savage, etc.) if people prefer the 75-25 risky urns to the 50-50 risky urn but prefer the 50-50 risky urn \textbf{A} above.

This experiment has the advantage of being simpler than our current experiment - it only has 3 possibilities instead of 101.

We could try also running the same experiment but with e.g. 60-40 and 40-60 in place of 75-25 and 25-75 above.  Try also e.g. 90-10 and 10-90.  See how extreme you have to make the asymmetry before people exhibit a preference for $RR$ over $AA$.

\newpage
\vspace*{\fill}
\begin{center}
    \section*{Online Appendix}
\end{center}
\vspace*{\fill}
\newpage

\section{Main Appendix}

\begin{itemize}
    \item Re-do analysis without removing attention screeners, 150, etc.
    \item Cognitive Uncertainty.
    \item 
\end{itemize}

\begin{table}[ht]
\centering
\scriptsize
\caption{\sc Contingent Variable Names}
\label{contvarnames}
\begin{tabular}{@{}lll@{}}
\toprule \toprule
\textbf{Name}  & \textbf{Definition}  &  \textbf{Description} \\[2pt] \midrule
$E^j$ & $K^j-U^j$ & CE difference in $j$-th elicitation of \textbf{Ellsberg} \\[2pt]
$Z^j$ & $K^j-UU^j$ & CE difference in $j$-th elicitation of \textbf{2Stage} \\[2pt]
$H^j$ & $K^j-C^j$ & CE difference in $j$-th elicitation of 50-50 vs. Halevy compound 50-50 \\[2pt]
$L^j$ & $K^j-CC^j$ & CE difference in $j$-th elicitation of 2-Stage simple 50-50 vs. compound 50-50 \\[2pt]
\Xhline{1.5\arrayrulewidth} \\[2pt]
$I^E$ & $\mathbbm{1}\{E^1 + E^2 > 0\}$ & Indicator for falling for classic Ellsberg paradox \\[2pt]
$I^T$ & $\mathbbm{1}\{T^1 + T^2 > 0\}$ & Indicator for falling for 2-Stage Ellsberg paradox \\[2pt]
$I^H$ & $\mathbbm{1}\{H^1 + H^2 > 0\}$ & Indicator for falling for Halevy paradox \\[2pt]
$I^L$ & $\mathbbm{1}\{L^1 + L^2 > 0\}$ & Indicator for falling for unambiguous 2-Stage paradox \\[2pt]
\Xhline{1.5\arrayrulewidth} \\[2pt]
 $F^{0-2}$ & $.5(UU^1/KK^1)+ .5 (UU^2/KK^2)$ & Ratio of certainty equivalents for $UU$ and $KK$ (averaged across 2 elicitations) \\[2pt]
$F^{0-3}$ & $UUU/3K$ & Ratio of certainty equivalents for $UUU$ and $KKK$ \\[2pt]
$F^{1-3}$ & $KUU/3K$ & Ratio of certainty equivalents for $KUU$ and $KKK$ \\[10pt]
\Xhline{1.5\arrayrulewidth} \\[2pt]
 $I^B$ & Treatment = $\mathcal{C}$ \& did ``Bounded U'' first & Indicator variable for having the "learning" section first  \\[2pt]
 $I^R$ & $R^2 =1$ and $R^3 = 0$ & Indicator variable for choosing the correct color in both practice questions \\[2pt]
$I^A$ & all $A^j = 1$ & Indicator variable for get all 3 attention screeners correct \\[2pt]\bottomrule \bottomrule
    \end{tabular}
    \label{tab:my_label}
\end{table}{}

\begin{table}[ht]
\centering
\tiny
\caption{\sc Decomposed Summary Statistics}
\begin{tabular}{lcccccccccccc}
\addlinespace
\toprule \toprule
\addlinespace
& &         $  T^1 $                    & $ T^2 $ & & $ E^1 $ & $ E^2 $ && $ H^1 $ & $ H^2 $ && $ L^1 $ & $ L^2 $ \\
\cmidrule{1-1}                      \cmidrule{3-4} \cmidrule{6-7} \cmidrule{9-10} \cmidrule{12-13}
Mean                                        & &                9.92        &           6.90          &&         13.49           &        10.79         & &            1.21          &           6.74        & &        11.31        &            3.13         \\
$ 95\% $ Conf. Interval & &     [  6.39 ,  13.44 ] &    [  3.43 ,  10.38 ]  &&    [  9.99 ,  17.00 ]  & [  7.22 ,  14.36 ]  & &    [ -7.10 ,   9.52 ]  &    [ -1.28 ,  14.76 ] & & [  2.73 ,  19.89 ] & [ -5.37 ,  11.64 ]  \\
\cmidrule{1-1}  \cmidrule{3-4} \cmidrule{6-7} \cmidrule{9-10} \cmidrule{12-13}
$ \rho $                            & &    \multicolumn{2}{c}{  0.152 }           &  &   \multicolumn{2}{c}{  0.287 }           & &          \multicolumn{2}{c}{  0.134 }              & &     \multicolumn{2}{c}{  0.221 }             \\
$ 1 - \rho $                        & &    \multicolumn{2}{c}{  0.848 }         &  &   \multicolumn{2}{c}{  0.713 }         & &          \multicolumn{2}{c}{  0.866 }            & &     \multicolumn{2}{c}{  0.779 }           \\
se $ 1 - \rho $             & &    \multicolumn{2}{c}{  0.033 }            &  &   \multicolumn{2}{c}{  0.032 }            & &          \multicolumn{2}{c}{  0.067 }               & &     \multicolumn{2}{c}{  0.066 }              \\
\addlinespace
\Xhline{1\arrayrulewidth}
$ N $                       & &    \multicolumn{2}{c}{ 880 }             &  &   \multicolumn{2}{c}{ 880 }             & &          \multicolumn{2}{c}{ 220 }                & &     \multicolumn{2}{c}{ 220 }              \\
\addlinespace
\bottomrule \bottomrule
\end{tabular}
\end{table}
\newpage


\section{Prolific Data Collection Details}

\paragraph{Fair attention check.}We did not use captcha for this reason ??? Instead we used attention checks. This has been developing these last years. However, amid those attention checks some are valid, some are not valid. Those not valid are..;Those valid, called ``fair attention checks'' are…We used these latter ones, following Prolific standards. 

\paragraph{Preventing duplicates.}Submissions to studies on Prolific are guaranteed to be unique by the firm\footnote{See Prolific unique submission guarantee policy \href{https://researcher-help.prolific.co/hc/en-gb/articles/360009220453-Preventing-participants-from-taking-your-study-multiple-times}{here}.}. Our system is set up such that each participant can have only one submission per study on Prolific. That is, each participant will be listed in your dashboard only once, and can only be paid once. On our side, we also prevent participants to take up several times our experiment in two steps. First, we enable the functionality ``Prevent Ballot Box Stuffing'' which permits to…Second we check participant ID and delete the second submission from the data set of the same ID if we find any. 

\paragraph{Drop-out rates.} Here put the drop out (or in the main text). 

\paragraph{High vs low-quality submissions.}Participants joining the Prolific pool receive a rate based on the quality of their engagement with the studies. If they are rejected from a study then they receive a malus. If they receive too much malus, then they are removed by the pool from the company\footnote{See Prolific pool removal Policy \href{https://researcher-help.prolific.co/hc/en-gb/articles/360009092394-Reviewing-submissions-How-do-I-decide-who-to-accept-reject-}{here}.}. Based on this long term contract, participants are incentivized to pay attention and follow the expectations of each study. Hence, a good research behavior has emerged on Prolific according to which, participants themselves can vol voluntarily withdraw their submissions if they feel they did a mistake such as rushing too much, letting the survey opened for a long period of time without engaging with it, and so on\footnote{See Prolific update regarding this behavior \href{https://researcher-help.prolific.co/hc/en-gb/articles/360009092394-Reviewing-submissions-How-do-I-decide-who-to-accept-reject-}{here}.}. According to these standards, we kept submissions rejections as low as possible, following standard in online experimental economics. Participants who fail at least one fair attention check are rejected and not paid. Following Prolific standards, participants who are statistical outliers (3 standard deviations below the mean) are excluded from the good complete data set. 

\paragraph{Payments and communication.}. We make sure to review participants’ submissions within within 24-48 hours after they have completed the study. This means that within this time frame, if we accept their submission, they receive their fixed and bonus payment. Otherwise, we reject their submissions and send to them a personalized e-mail(\footnote{Partially-anonymized through Prolific messaging app which put the researcher’s name visible to the participants and only the participantsID visible to the researcher.}), detailing the reason of the rejection, leaving participants the opportunity to contact us afterwards if they firmly believe the decision to be unfair (motivate their perspective). Participants can also contact us at any time if they encounter problems with our study or just have questions about it.

\section{Variables Dictionary}
\subsection{Independent Variables}
\begin{table}[ht]
\scriptsize
\centering
\caption*{\sc independent variable names}
    \begin{tabular}{@{}lll@{}}
     \toprule \toprule
\textbf{Stata/Paper}  & \textbf{Data File}  &  \textbf{Elicitation Description} \\[2pt] \midrule
$K^1$ & Balc1a & 1st elicitation of risk preferences in one-stage Ellsberg\\
$K^2$ & Final1a & 2nd  elicitation of risk preferences in one-stage Ellsberg \\[1pt] \Xhline{0.2\arrayrulewidth}
$U^1$ & Balc1d & 1st  elicitation of ambiguous preferences in one-stage Ellsberg\\
$U^2$ & Final1c & 2nd  elicitation of ambiguous preferences in one-stage Ellsberg \\[1pt] \Xhline{1.5\arrayrulewidth}
$KK^1$ & Balu1a & 1st elicitation of risk preferences in two-stage Ellsberg\\
$KK^2$ & Matu1a & 2nd elicitation of risk preferences in two-stage Ellsberg\\[4pt] \Xhline{0.2\arrayrulewidth}
$UU^1$ & Balu1b & 1st elicitation of ambiguous preferences in two-stage Ellsberg\\
$UU^2$ & Matu1b & 2nd elicitation of ambiguous preferences in two-stage Ellsberg\\[4pt] \Xhline{0.2\arrayrulewidth}
$UK^1$ & Balu1c & \\
$UK^2$ & Matu1c & \\[4pt] \Xhline{0.2\arrayrulewidth}
$KU^1$ & Balu1d & \\
$KU^2$ & Matu1d & \\[4pt] \Xhline{1.5\arrayrulewidth}
$KKK$ & Balu2a & elicitation of risk preferences in 3-stage Ellsberg \\
$UUU$ & Balu2b &  elicitation of ambiguous preferences in 3-stage Ellsberg \\
$KUU$ & Balu2c & \\[4pt] \Xhline{1.5\arrayrulewidth}
$II$ & Balu4 & 2-Stage gamble with indepedent ambiguous urns \\[4pt]  \Xhline{1.5\arrayrulewidth}
$C^1$ & Lotte1 & 1st Halevy compound 50-50 lottery \\
$C^2$ & Final2a & \\[4pt] \Xhline{1.5\arrayrulewidth}
$CC^1$ & Lotte2 & 1st 2-stage Halevy \\
$CC^2$ & Final2b & \\[4pt] \Xhline{1.5\arrayrulewidth}
$BB^{40-60}$ & Cmu1b & 2-stage Ellsberg with bounded U ($40 \leq R \leq 60$) \\
$BB^{60-100}$ & Cmu2b & 2-stage Ellsberg with bounded U ($60 \leq R \leq 100$)   \\
$BB^{95-100}$ & Cmu4b & 2-stage Ellsberg with bounded U ($95 \leq R \leq 100$)  \\[2pt] 
$R^1$ & Answered "red" on Mp1 &  \\[2pt] 
$R^2$ & Answered "red" on Mp2 & Picked the CORRECT color in practice question 2  \\[2pt] 
$R^3$ & Answered "red" on Mp3 & Picked the WRONG color in practice question 3  \\[2pt] 
$P^1$ & $Q78$ & Indicator variable for get $P^1 = 1$, i.e., correct := ``32 Blue balls and 95 Red balls'' \\[2pt] 
$P^2$ & $Q1777$ &  Indicator variable for get $P^2=1$, i.e., correct := ``2'' \\[2pt] 
$P^3$ & $Q80$ &   Indicator variable for get $P^3=1$, i.e., correct := ``\$1'' \\[2pt]
$A^1$ & $Q13$ & Indicator variable for get $A^1=1$, i.e., correct := ``orange'' \\[2pt] 
$A^2$ & $Q22$ & Indicator variable for get $A^2=1$, i.e., correct := ``11'' \\[2pt] 
$A^3$ & $Q30$ & Indicator variable for get $A^3=1$, i.e., correct := ``blue'' \\[2pt] 
\bottomrule \bottomrule
    \end{tabular}
    \label{tab:my_label}
\end{table}{}

\subsection{Dependent Variables}

\begin{table}[ht]
\centering
\scriptsize
\caption*{\sc dependent variable names}
\begin{tabular}{@{}lll@{}}
\toprule \toprule
\textbf{Stata/Paper}  & \textbf{Definition}  &  \textbf{ Description} \\[2pt] \midrule
$E^j$ & $K^j-U^j$ & Certain equivalent difference in $j$-th elicitation of 1-stage Ellsberg \\[2pt]
$T^j$ & $KK^j-UU^j$ & Certain equivalent difference in $j$-th elicitation of 2-stage Ellsberg \\[2pt]
$H^j$ & $K^j-C^j$ & Certain equivalent difference in $j$-th elicitation of 50-50 vs. Halevy compound 50-50 \\[2pt]
$L^j$ & $KK^j-CC^j$ & Certain equivalent difference in $j$-th elicitation of $KK$ vs. $CC$ \\[10pt]
\Xhline{1.5\arrayrulewidth} \\[2pt]
 $F^{0-2}$ & $.5(UU^1/KK^1)+ .5 (UU^2/KK^2)$ & Ratio of certainty equivalents for $UU$ and $KK$ (averaged across 2 elicitations) \\[2pt]
$F^{0-3}$ & $UUU/KKK$ & Ratio of certainty equivalents for $UUU$ and $KKK$ \\[2pt]
$F^{1-3}$ & $KUU/KKK$ & Ratio of certainty equivalents for $KUU$ and $KKK$ \\[10pt]
\Xhline{1.5\arrayrulewidth} \\[2pt]
$I^E$ & $E^1 + E^2 > 0$ & Indicator variable for having a larger Certain equivalent for $K$ than $U$ \\[2pt]
$I^T$ & $T^1 + T^2 > 0$ & Indicator variable for having a larger Certain equivalent for $KK$ than $UU$ \\[2pt]
$I^H$ & $H^1 + H^2 > 0$ & Indicator variable for having a larger Certain equivalent for $K$ than $C$ \\[2pt]
$I^L$ & $L^1 + L^2 > 0$ & Indicator variable for having a larger Certain equivalent for $KK$ than $CC$ \\[2pt]
 $I^B$ & Treatment = $\mathcal{C}$ \& did ``Bounded U'' first & Indicator variable for having the "learning" section first  \\[2pt]
 $I^R$ & $R^2 =1$ and $R^3 = 0$ & Indicator variable for choosing the correct color in both practice questions \\[2pt]
$I^A$ & all $A^j = 1$ & Indicator variable for get all 3 attention screeners correct \\[2pt]\bottomrule \bottomrule
    \end{tabular}
    \label{tab:my_label}
\end{table}{}
Note on the naming convention for first few items: $E$=\textbf{E}llsberg, $T$=\textbf{T}wo-stage, $H$=\textbf{H}alevy, $L$ = compound \textbf{L}ottery

\newpage
\vspace*{\fill}
\begin{center}
    \section*{MPL Example}
    Complete experimental instructions available online
\end{center}
\vspace*{\fill}

\includegraphics[scale=0.55]{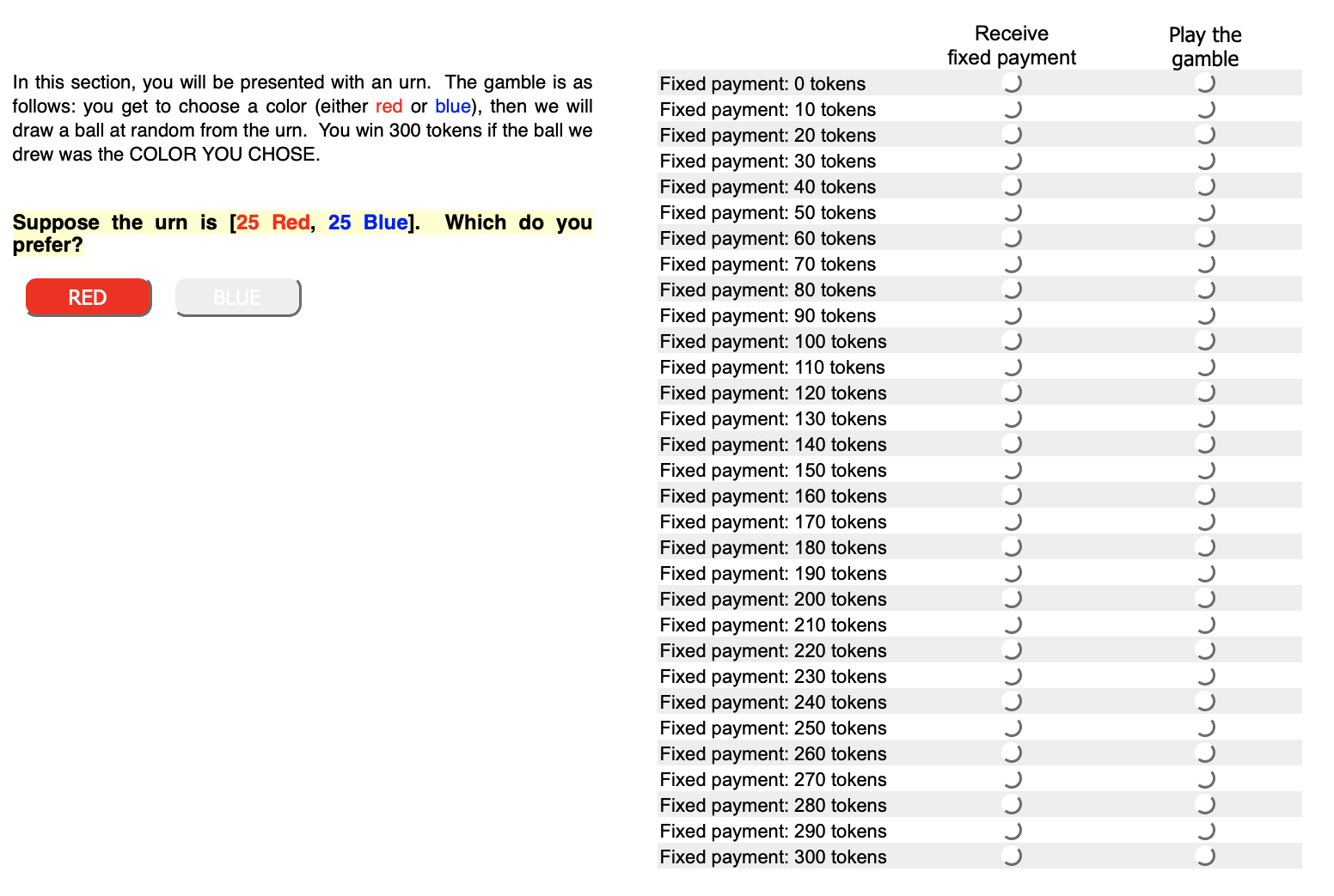} \vspace*{\fill} \newpage

\end{document}